\documentclass[10pt]{article}
\usepackage[margin=0.8in]{geometry}
\usepackage{authblk}
\usepackage{sectsty}
\usepackage{amsmath}
\usepackage{graphicx}
\usepackage{enumerate}
\usepackage{natbib}
\usepackage{url}
\usepackage{hyperref}
\usepackage{amsfonts,amssymb}
\usepackage{subfig,etoolbox}
\sectionfont{\large}
\subsectionfont{\large}
\numberwithin{equation}{section}
\newtheorem{def1}{Definition}
\numberwithin{def1}{section}
\newtheorem{theo1}{Theorem}
\numberwithin{theo1}{section}
\newtheorem{cor1}{Corollary}
\numberwithin{cor1}{section}

\newcommand\blfootnote[1]{%
  \begingroup
  \renewcommand\thefootnote{}\footnote{#1}%
  \addtocounter{footnote}{-1}%
  \endgroup}
\title{\fontsize{15}{18}\selectfont \textbf{Finite mixture copulas for modeling dependence \\ in longitudinal count data}}
\author[]{\fontsize{10}{12}\selectfont Subhajit Chattopadhyay}
\affil[]{\small Bhubandanga, Bidya Sagar Path, Bolpur, West Bengal, India, PIN: 731204}
\date{}
\begin{document}
\bigskip
\maketitle
\begin{abstract}
Dependence modeling of multivariate count data has garnered significant attention in recent years. Multivariate elliptical copulas are typically preferred in statistical literature to analyze dependence between repeated measurements of longitudinal data since they allow for different choices of the correlation structure. But these copulas lack in flexibility to model dependence and inference is only feasible under parametric restrictions. In this article, we propose employing finite mixtures of elliptical copulas to better capture the intricate and hidden temporal dependencies present in discrete longitudinal data. Our approach allows for the utilization of different correlation matrices within each component of the mixture copula. We theoretically explore the dependence properties of finite mixtures of copulas before employing them to construct regression models for count longitudinal data. Inference for this proposed class of models is based on a composite likelihood approach, and we evaluate the finite sample performance of parameter estimates through extensive simulation studies. To validate our models, we extend traditional techniques and introduce the $t$-plot method to accommodate finite mixtures of elliptical copulas. Finally, we apply our models to analyze the temporal dependence within two real-world longitudinal datasets and demonstrate their superiority over standard elliptical copulas.
\end{abstract}
\blfootnote{\textbf{Correspondence}: Subhajit Chattopadhyay, email: subhajit@marketopper.com}
\noindent
{\textbf{Keywords:} composite likelihood; count responses; longitudinal data; mixture copula; temporal dependence.}
\section{Introduction}
In longitudinal studies for count data, a small number of repeated count responses along with a set of multidimensional covariates are collected from a large number of independent individuals. Two main aspects of analyzing such data is to find the effects of the covariates on the repeated count responses and to measure the strength of dependence between them across time. \citet{weiss2005modeling} or \citet{sutradhar2011dynamic} are some of the excellent books which provide with concepts and ideas for developing correlation models for discrete longitudinal data. \citet{gibbons2010advances} reviewed recent developments of discrete longitudinal data analysis. In the context of count data, some of the seminal works includes \citet{thall1990some} and \citet{jowaheer2002analysing}, who considered mixed effect approaches for modeling longitudinal count data. \citet{sutradhar2003overview} provided different estimation techniques for count regression models based on generalized estimating equations. But these mixed model approaches comes with computational complexities due to the involvement of multidimensional random effects.
\par The interest in the copula framework is steadily rising among statisticians, particularly for modeling non-Gaussian count longitudinal data. Copulas offer exceptional flexibility in modeling multivariate distributions and naturally interpret the dependence between random variables. This approach allows for the introduction of various dependence structures into regression models while permitting arbitrary selection of marginals. \citet{nikoloulopoulos2009modeling} and \citet{nikoloulopoulos2010regression} were among the first to demonstrate successful applications of copulas to multivariate count data. They highlighted that while some desirable properties of copulas may not hold directly when applied to discrete data, modeling remains feasible due to Sklar's theorem, which uniquely identifies copulas over the product range of marginals. Elliptical copulas, recommended by \citet{sun2008heavy}, are often employed to model non-Gaussian longitudinal data, as they account for within-subject dependencies via correlation matrices, thus capturing time series behavior. \citet{shi2012multivariate} and \citet{shi2014longitudinal} extended this approach to model longitudinal insurance claim data. However, the parametric restrictions of these copulas limit their ability to capture complex dependence patterns effectively. To overcome the limitations of standard multivariate copulas, several authors have proposed the use of vine copulas (see, \citet{aas2009pair}). \citet{smith2010modeling} incorporated pair-copula constructions to model joint dependence of longitudinal data but without considering covariates. Nevertheless, vine copulas can involve numerous parameters to estimate, posing computational challenges in various scenarios. Recent research has combined copulas with finite mixture models to better understand complex dependence patterns and increase model flexibility. Finite mixture models, which represent probabilistic models as weighted sums of a few parametric densities, are instrumental in uncovering hidden structures in data. For instance, \citet{yang2022time} employed a time-varying mixture copula model in a semiparametric setup to analyze the time series behavior of stock market data.
\par In this article we focus our attention to model the unknown correlation structure of longitudinal count data. Our motivation comes from the fact that in a mixture of elliptical copulas, different correlation structures can be used in each component copulas. The magnitudes of the weights in a mixture copula model signify the importance of the corresponding component copulas. Moreover, the model parameters can be efficiently estimated using the composite likelihood method followed by a two stage procedure. The contribution to this manuscript is two fold. Firstly, we explicitly derive the dependence properties of mixture copula models under both continuous and discrete setup. As far as we reviewed the literature, mixture copulas have been applied to discrete data but their dependence properties under the discrete setup were not explicitly derived. Recently \citet{safari2020modelling} introduced population version of Spearman's rho for discrete data and applied some well known Archimedean copulas to model bivariate count data. In our applications, we utilize regression models for longitudinal count data, where temporal dependencies are captured using a mixture of elliptical copulas, providing deeper insights into the underlying dependence dynamics. 
\par The remainder of the article is structured as follows. Section \ref{sec2} provides a review of standard definitions of mixture and elliptical copulas. In Section \ref{sec3}, we derive theoretical properties such as Kendall's tau and Spearman's rho for a general mixture copula under both continuous and discrete cases. Section \ref{sec4} outlines standard marginal regression models commonly used for modeling longitudinal count data. We discuss the two-stage composite likelihood estimation method for mixture copula-based count regression models in Section \ref{sec5} and we cover standard model validation techniques, which extends the $t$-plot method of model validation for finite mixtures of elliptical copulas. Rigorous simulation studies are conducted in Section \ref{sec7} to assess the finite sample performance of our proposed class of models under different sample sizes. In Section \ref{sec8}, we apply our methods to model the temporal dependence of two real-world datasets, comparing the fits of mixture copulas with standard elliptical copulas and demonstrating substantial improvements. Utilizing the derived expressions of Kendall's tau and Spearman's rho, we estimate the concordance matrices of these datasets and show their close alignment with empirical versions. Finally, we conclude this article with a general discussion in Section \ref{sec9}.
\section{The K-finite mixture of multivariate copulas}\label{sec2}
Mixture models have been widely studied in the statistical literature; see for example the books by \citet{titterington1985statistical}; \citet{bohning1999computer} and \citet{fruhwirth2006finite}, among others. \citet{mclachlan2019finite} provided with the basics of mixture modeling and its applications. As more flexible multivariate distributions can be obtained by mixing different distributions of the same dimension (might not be from the same family), similarly one can use mixture of different copulas to introduce different dependence characteristics in a statistical model. In many real-life applications, a single parametric copula might be insufficient to capture all important features when performing analysis. Finite mixture copula models have been previously studied in the literature (see, \citet{arakelian2014clustering}; \citet{kosmidis2016model} or \citet{zhuang2022bayesian}).
\par A K-finite mixture copula is defined as
\begin{equation}\label{defmix1}
C_{\text{mix},d} = \sum_{l=1}^K \pi_l C_{l,d}(.|\mathbf{\phi}_l), \quad \sum_{l=1}^K \pi_l = 1, \pi_l \geq 0, \quad \forall \; l = 1,\dots,K  
\end{equation}
where $C_{l,d}(.|\mathbf{\phi}_l)$ denotes a single $d$-dimensional multivariate copula component which has a mixture weight $\pi_l$, and $K$ is a pre-defined hyperparameter. It is straight forward to check that the distribution function in (\ref{defmix1}) is a copula. The density function of the mixture copula can be simply obtained as
\begin{equation}\label{mixden1}
c_{\text{mix},d}(u_1,\dots,u_d|\mathbf{\eta}) = \sum_{l=1}^K \pi_l c_{l,d}(u_1,\dots,u_d|\mathbf{\phi}_l) 
\end{equation}
where $\mathbf{\eta} = \{\pi_l,\mathbf{\phi}_l; l = 1,\dots,K\}$ denotes all dependence parameters, $\mathbf{\phi}_l$ contains the copula parameters of the $l$-th component and $\mathbf{u}$ is the uniform vector. The choice of the copula components in the mixture (\ref{defmix1}) is arbitrary, but in this article we restrict to the multivariate elliptical copulas.
\subsection{Elliptical copulas}
Multivariate copulas generated from elliptical distributions as Gaussian or Student-$t$ are popular in the literature of statistics and econometrics for their simplicity in terms of parametric inference (see, \citet{xue2000multivariate}; \citet{demarta2005t}). \citet{frahm2003elliptical} discussed the dependence structures generated by elliptical distributions and their limitations. We proceed with some standard definitions as follows.
\begin{def1}
A $d$-dimensional copula is said to be a Gaussian copula if
\begin{equation}
C_d(\mathbf{u}|\mathbf{\Sigma}) = \Phi_d(\Phi_1^{-1}(u_1),\dots,\Phi_1^{-1}(u_d)|\mathbf{\Sigma})
\end{equation}
where $\Phi_1^{-1}(u_j)$ denotes the inverse of the CDF of $X_j \sim N_1(0,1)$ distribution and $\mathbf{\Sigma}$ is a correlation matrix.
\end{def1}
\begin{def1}
A $d$-dimensional copula is said to be a Student-$t$ copula if
\begin{equation}
C_d(\mathbf{u}|\mathbf{\Sigma},\mathbf{\nu}) = T_d(T_1^{-1}(u_1|\mathbf{\nu}),\dots,T_1^{-1}(u_d|\mathbf{\nu})|\mathbf{\Sigma},\mathbf{\nu})
\end{equation}
where $T_1^{-1}(u_j|\mathbf{\nu})$ denotes the inverse of the CDF of $T_j \sim T_1(0,1,\mathbf{\nu})$ distribution and $\mathbf{\Sigma}$ is a correlation matrix.
\end{def1}
Student-$t$ copula has an additional degrees of freedom parameter $\mathbf{\nu}$, which accounts for possible tail dependence in the data. We consider the mixture of Gaussian and Student-$t$ copulas for modeling the temporal dependence of longitudinal count data. As \citet{sun2008heavy} emphasized, elliptical copulas are more useful when the dimension of the data is moderate to high since all lower dimensional sub-copulas stay in the same parametric family (\citet{frees2006copula}).
\subsection{Model identifiability}
Identifiability of finite mixture models is considered to be a critical aspect in statistical analysis. Identifiability of parametric mixture densities have been studied in the seminal works by \citet{teicher1963identifiability} and \citet{yakowitz1968identifiability}. They pointed out that most finite mixtures of continuous densities are identifiable; an exception is a mixture of uniform densities.  In this paper we implement copulas to discrete margins which poses identifiability issues as one could use different copulas to construct the same discrete probability distribution. \citet{genest2007primer} highlighted the consequences in inference for the lack of uniqueness of the copula under discrete margins. Hence, it is reasonable to proceed by using one particular choice of copulas in applications. \citet{holzmann2006identifiability} discussed some theoretical results on identifiability of mixture models constructed from elliptical distributions. They showed that elliptical mixtures are identifiable. \citet{liu2023mixture} argued that finite mixture of Gaussian copulas is identifiable followed from the fact that the quantiles have the distribution of a multivariate normal mixture. Therefore, in this paper we only consider the mixture of elliptical copulas within their own parametric family that is mixture with mixing components belonging to the same family of distributions. Another reason for that is they are easy to simulate and have simpler parametric inference using composite likelihood which will be discussed latter on. While it is challenging to validate the identifiability of the proposed class models within our framework, identifiability is not a concern in our two-step method. This is because, in the second step, all mixture copulas, which are theoretically identifiable, are estimated.
\section{Dependence properties}\label{sec3}
Here we provide a comprehensive theoretical analysis of mixture copula models for both continuous and discrete margins. The results presented in this section are for a general mixture copula model. For continuous random variables, dependence as measured by Kendall's tau or Spearman's rho is associated only with the copula parameters (see, \cite{nelsen2006introduction}). Therefore the dependence parameters $\mathbf{\eta}$, of the mixture copulas in (\ref{mixden1}) can be interpreted with respect to Kendall's tau and Spearman's rho, which are bounded in the interval $[-1,1]$. For notational simplicity in this section we consider $C_{\text{mix}}$ to be of dimension $d = 2$, that is for the bivariate copulas. First we derive the population versions of Kendall's tau and Spearman's rho for continuous random variables as follows.
\begin{theo1}\label{thm1}
Let $(X_1,X_2)^\intercal$ be a bivariate continuous random vector having dependence of finite mixture copula $C_{\text{mix}}$ defined in (\ref{defmix1}), with marginal distribution functions $F_j, j = 1,2$. The population version of Kendall's tau for $X_1$ and $X_2$ is given by
\begin{equation}\label{taucon1}
\tau(C_{\text{mix}}) = \sum_{l=1}^K \pi^2_l\tau(C_l) + 2\sum_{l<m}^K \pi_l\pi_m Q_{lm},
\end{equation}
where $Q_{lm} = Q(C_l,C_m)$ is the concordance function defined for copulas $C_l$ and $C_m$ and $\tau(C_l)$ is the Kendall's tau corresponding to copula $C_l$, for $l = 1,\dots,K$. 
\end{theo1}
Note that the concordance function is symmetric in its arguments for a continuous random vector (\citet{nelsen2006introduction}; \citet{arakelian2014clustering}), but this doesn't hold for the discrete case.
\begin{cor1}
For mixture of Gaussian and Student-$t$ copula, the population version of Kendall's tau can be obtained in a closed form expression as -
\begin{equation}\label{kendtau2}
\tau(C_{\text{mix}}) = \frac{2}{\pi} \sum_{l=1}^K \pi^2_l \arcsin(\rho_l) + \frac{4}{\pi} \sum_{l<m}^K \pi_l\pi_m \arcsin\Big(\frac{\rho_l + \rho_m}{2}\Big),
\end{equation}
where $\rho_l$ is the correlation parameter of bivariate Gaussian or Student-$t$ copula, for $l = 1,\dots,K$.
\end{cor1}
\begin{theo1}\label{thm2}
Let $(X_1,X_2)^\intercal$ be a bivariate continuous random vector having dependence of finite mixture copula $C_{\text{mix}}$ defined in (\ref{defmix1}), with marginal distribution functions $F_j, j = 1,2$. The population version of Spearman's rho for $X_1$ and $X_2$ is given by
\begin{equation}\label{rhocon1}
\rho(C_{\text{mix}}) = \sum_{l=1}^K \pi_l\rho(C_l),
\end{equation}
where $\rho(C_l)$ is the Spearman's rho corresponding to copula $C_l$, for $l = 1,\dots,K$.
\end{theo1}
Expression of Spearman's rho for Gaussian mixture copula can be obtained in a closed form, but for Student-$t$ no closed form is available (see, \citet{joe2014dependence}).
\par
\begin{figure}
    \centering
    \includegraphics[width = 12cm]{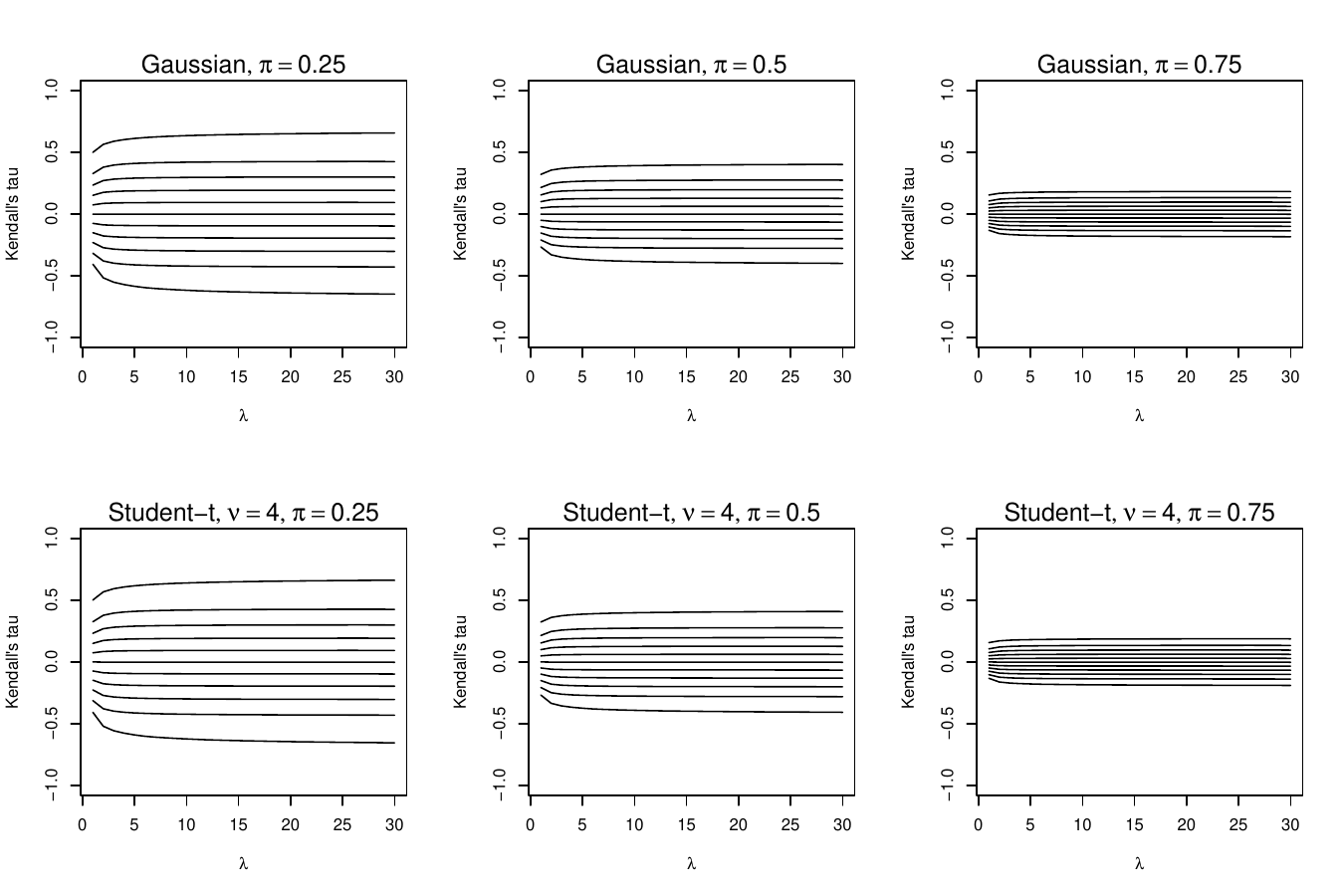}
    \caption{Kendall's tau values computed using Gaussian and Student-$t$ ($\nu = 4$) mixture copulas with different mixing proportions and Poisson marginal distributions with the same location parameter $\mu = 1,\dots,30$. Higher curves corresponding to higher values of the copula parameter.}
    \label{fig:Kendall1}
\end{figure}

\begin{figure}
    \centering
    \includegraphics[width = 12cm]{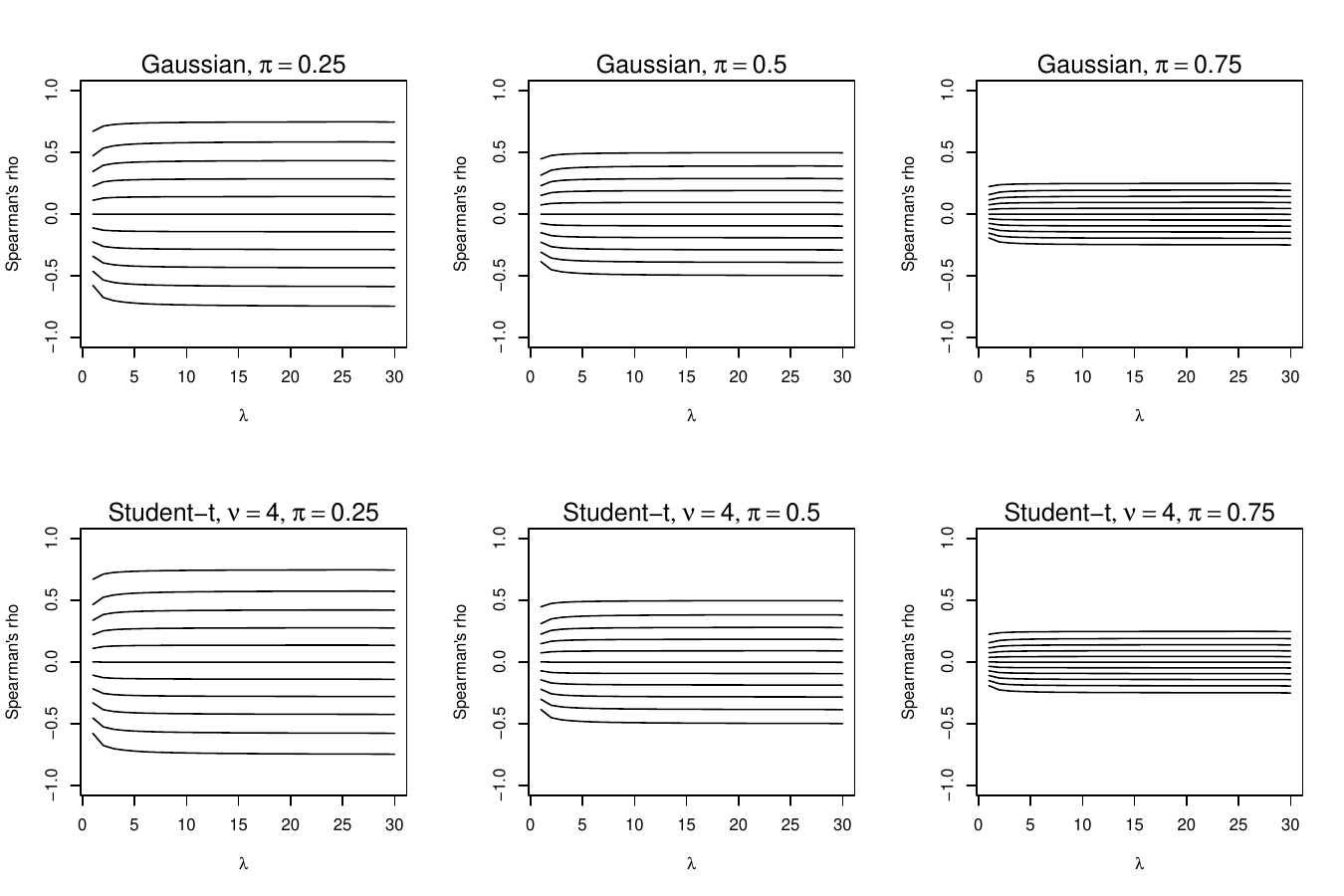}
    \caption{Spearman's rho values computed using Gaussian and Student-$t$ ($\nu = 4$) mixture copulas with different mixing proportions and Poisson marginal distributions with the same location parameter $\mu = 1,\dots,30$. Higher curves corresponding to higher values of the copula parameter.}
    \label{fig:Spearman1}
\end{figure}

\begin{figure}
    \centering
    \includegraphics[width = 12cm]{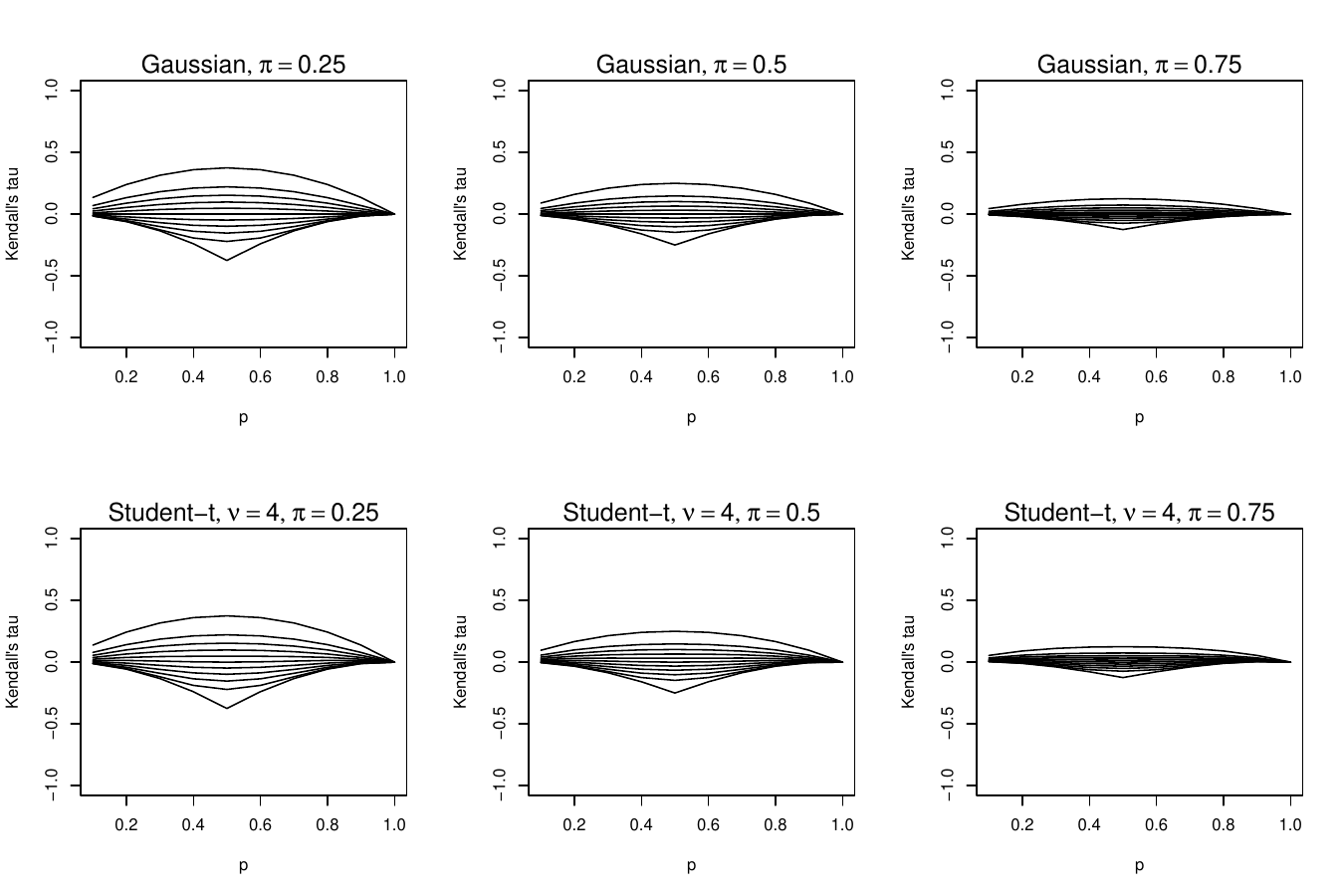}
    \caption{Kendall's tau values computed using Gaussian and Student-$t$ ($\nu = 4$) mixture copulas with different mixing proportions and Bernoulli marginal distributions with the same location parameter $p = 0.1,\dots,1.0$. Higher curves corresponding to higher values of the copula parameter.}
    \label{fig:Kendall2}
\end{figure}

\begin{figure}
    \centering
    \includegraphics[width = 12cm]{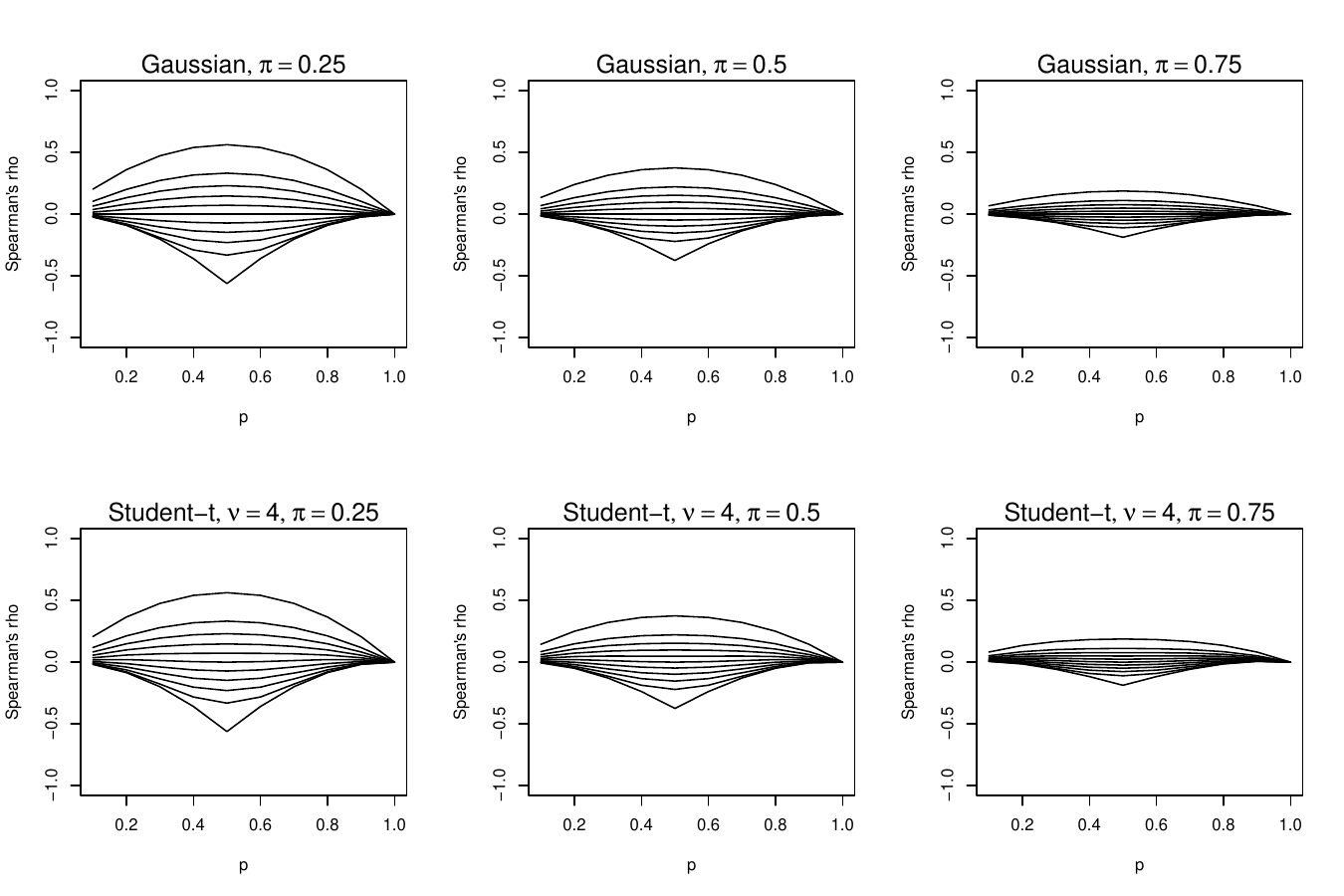}
    \caption{Spearman's rho values computed using Gaussian and Student-$t$ ($\nu = 4$) mixture copulas with different mixing proportions and Bernoulli marginal distributions with the same location parameter $p = 0.1,\dots,1.0$. Higher curves corresponding to higher values of the copula parameter.}
    \label{fig:Spearman2}
\end{figure}

 When the marginal distributions are discrete, these concordance based measures depends on the marginal distributions as well as the copula. \citet{denuit2005constraints} and \citet{mesfioui2005properties} studied the population version of Kendall's tau applied to discrete data, which is not distribution free and has a range narrower than $[-1,1]$. \citet{nikoloulopoulos2010regression} previously derived the population version of Kendall's tau under discrete marginals. Recently, \citet{safari2020modelling} derived the population version of Spearman's rho followed by continuous extension of discrete random variables. Followed from these, we derive the population versions of these concordance measures of mixture of copulas for the discrete case. Note that when the marginal distributions are discrete, the probability of tie is positive which needs to be considered in the derivation.
\begin{theo1}\label{thm3}
Let $(X_1,X_2)^\intercal$ be a bivariate integer valued random vector having dependence of finite mixture copula $C_{\text{mix}}$ defined in (\ref{defmix1}), with marginal distribution functions $F_j, j = 1,2$ and mass functions $f_j, j = 1,2$. The population version of Kendall's tau for $X_1$ and $X_2$ is given by
\begin{equation}\label{taudis1}
\tau^*(C_{\text{mix}}) = \sum_{l=1}^K \pi^2_l\tau^*(C_l) + \sum_{l\neq m}^K \pi_l\pi_m Q^*_{lm},
\end{equation}
\begin{align}
\text{where} \quad \tau^*(C_l) & = \sum_{x_1=0}^\infty \sum_{x_2=0}^\infty h_l(x_1,x_2)\{4C_l(F_1(x_1-1),F_2(x_2-1)|\mathbf{\phi}_l) - h_l(x_1,x_2)\} \nonumber \\
& + \sum_{x_1=0}^\infty f_1^2(x_1) + \sum_{x_2=0}^\infty f_2^2(x_2) - 1, \\
Q^*_{lm} & = \sum_{x_1=0}^\infty \sum_{x_2=0}^\infty h_m(x_1,x_2)\{4C_l(F_1(x_1-1),F_2(x_2-1)|\mathbf{\phi}_l) - h_l(x_1,x_2)\} \nonumber \\
& + \sum_{x_1=0}^\infty f_1^2(x_1) + \sum_{x_2=0}^\infty f_2^2(x_2) - 1,
\end{align}
\begin{align}\label{hval1}
\text{and} \quad h_l(x_1,x_2) & = C_l(F_1(x_1),F_2(x_2)|\mathbf{\phi}_l) - C_l(F_1(x_1-1),F_2(x_2)|\mathbf{\phi}_l) \nonumber \\ & - C_l(F_1(x_1),F_2(x_2-1)|\mathbf{\phi}_l) + C_l(F_1(x_1-1),F_2(x_2-1)|\mathbf{\phi}_l).
\end{align}
\end{theo1}
\begin{theo1}\label{thm4}
Let $(X_1,X_2)^\intercal$ be a bivariate integer valued random vector having dependence of finite mixture copula $C_{\text{mix}}$ defined in (\ref{defmix1}), with marginal distribution functions $F_j, j = 1,2$ and mass functions $f_j, j = 1,2$. The population version of Spearman's rho for $X_1$ and $X_2$ is given by
\begin{equation}\label{rhodis1}
\rho^*(C_{\text{mix}}) = \sum_{l=1}^K \pi_l\rho^*(C_l),
\end{equation}
\begin{align}
\text{where} \quad \rho^*(C_l) & = \sum_{x_1=0}^\infty \sum_{x_2=0}^\infty h_l(x_1,x_2)\{6F_1(x_1-1)F_2(x_2-1) + 6(1-F_1(x_1))(1-F_2(x_2)) \nonumber \\ & -3f_1(x_1)f_2(x_2)\} + 3(\sum_{x_1=0}^\infty f_1^2(x_1) + \sum_{x_2=0}^\infty f_2^2(x_2) - 1)
\end{align}
and $h_l(x_1,x_2)$ is the joint probability mass function same as defined in (\ref{hval1}).
\end{theo1}
It is evident that the Spearman's rho of convex combination of copulas equals the convex combination of the individual Spearman's rho for both continuous and discrete case. The derived results help us to clarify that in the discrete case the marginals do affect the dependence measures. Based on the expressions in (\ref{taucon1}), (\ref{rhocon1}), (\ref{taudis1}) and (\ref{rhodis1}) one can see that even if the component copulas imply independence, the resulting mixture may imply dependence. In Figures \ref{fig:Kendall1}, \ref{fig:Spearman1}, \ref{fig:Kendall2} and \ref{fig:Spearman2} we provide the plotted values of Kendall's tau and Spearman's rho for different elliptical mixture copulas and different marginal distributions. We consider a $2$-component mixture with the first copula set to independence ($\rho = 0$) and the second copula's correlation parameter varied over $[-1,1]$. For Poisson marginals with the same parameter, the association with Kendall's tau and Spearman's rho becomes negligible for values greater than 10. For Bernoulli marginals with the same parameter, the association is minimal when the success probability $p = 0.5$. Elliptical mixture copulas are symmetric and allow both positive and negative dependence. As the proportion of the independence copula increases, Kendall's tau and Spearman's rho approach zero. Lastly we need to discuss the tail dependence of a mixture copula. Tail dependence quantifies the degree of dependence in the joint lower or joint lower of a multivariate distribution. Here we consider the bivariate tail dependence only, but there are multivariate extensions to the concept in the literature (see, \citet{jaworski2010copula}). The following Theorem states the tail behavior of bivariate mixture copulas.
\begin{theo1}\label{thm5}
Let $\lambda_U(C_l)$ and $\lambda_L(C_L)$ be the tail dependence coefficients of the component copula $C_l$, provided these exists and $\pi_l$ be the mixing proportion for $l = 1,\dots,K$. Then the upper and lower tail dependence coefficients are given as
\begin{equation}\label{mixtailu1}
\lambda_U(C_{\text{mix}}) = \sum_{l=1}^K \pi_l \lambda_U(C_l) \quad \text{and} \quad \lambda_L(C_{\text{mix}}) = \sum_{l=1}^K \pi_l \lambda_L(C_l), \quad \text{respectively}.     
\end{equation}
\end{theo1}
It is direct that mixture of Gaussian copulas has zero tail dependence. For mixture of bivariate Student-$t$, both of the tail dependence coefficients are same (\citet{demarta2005t}), i.e.
\begin{equation}\label{bivtmix1}
\lambda(C_{\text{mix}}) = 2\sum_{l=1}^K \pi_l T\Bigg(-\frac{\sqrt{(\mathbf{\nu} + 1)(1 - \rho_l)}}{\sqrt{1 + \rho_l}}\Bigg|\mathbf{\nu}\Bigg).
\end{equation}
All proofs related to this Section are provided in Appendix \ref{appndx1}.
\section{Modeling longitudinal count data}\label{sec4}
From now on we specifically focus on the applications of mixture copulas by considering regression models with covariates. Suppose that $\mathbf{Y_i} = (Y_{i1},\dots,Y_{in_i})^\intercal$ be a non-negative integer valued dependent random vector for the $i$-th subject, where $Y_{ij}$ denotes the observation at time $t_{ij}$. Let $\mathbf{x}_{ij}$ be a $p$-dimensional vector of covariates observed for the $i$-th subject at time $t_{ij}$ and $\mathbf{\beta}$ is the $p \times 1$ vector of regression parameters. In the longitudinal set-up, the components of the vector $\mathbf{Y}_i$ are repeated responses, which are likely to be correlated. Since for discrete random variables the copula is uniquely identified in in the product range of the marginals, we assume restrictively that the marginal models for the repeated responses are correctly specified. Here we proceed by applying two most commonly used distributions for count responses. Poisson regression assumes that the repeated count responses marginally follows a Poisson distribution with CDF
\begin{equation}
F_{ij}(y_{ij}|\mathbf{x}_{ij},\mathbf{\beta}) = \sum_{k=0}^{y_{ij}} \frac{e^{-\mu_{ij}}\mu_{ij}^k}{k!}, \quad j = 1,\dots,n_i,
\end{equation}
where $E(Y_{ij}) = Var(Y_{ij}) = \mu_{ij} = \exp(\mathbf{x}_{ij}\mathbf{\beta})$. Since Poisson distribution doesn't allow for overdispersion in longitudinal count data, we also consider Negative Binomial regression model for the count responses. The CDF of a Negative Binomial marginal can be written as
\begin{equation}
F_{ij}(y_{ij}|\mathbf{x}_{ij},\mathbf{\beta}) = \sum_{k=0}^{y_{ij}} \left(\begin{array}{c} \psi + k - 1 \\ k \end{array}\right) \frac{\psi^\psi \mu_{ij}^k}{(\psi + \mu_{ij})^{\psi + k}}, \quad j = 1,\dots,n_i, 
\end{equation}
where $E(Y_{ij}) = \mu_{ij} = \exp(\mathbf{x}_{ij}\mathbf{\beta})$ and $Var(Y_{ij}) = \mu_{ij} + \mu_{ij}^2/\psi$. Point to be noted that both of these distributions are defined on same support of non-negative integers. 
\par Since in our approach we employ mixture of elliptical copulas it allows to use different correlation structures into each copula component. This can be nicely interpreted as we can show which correlation structure is present in the data set in what proportion. Common choices for modeling serial dependence in longitudinal data are auto-regressive (AR(1)), moving average (MA(1)) or exchangeable (EX) correlation structures. Under the AR(1) structure, the correlation between the errors on a subject decline exponentially with the distance between the observations. In terms of the correlation matrix $\mathbf{\Sigma}$ in the elliptical copulas, it can be expressed as
\begin{equation}
\rho_{jk} = \exp(-\xi|t_j - t_k|), \xi > 0, \quad 1 \leq j < k \leq n_i,
\end{equation}
where $\xi > 0$ is the auto-regressive correlation parameter, where $\rho_{j,k}$'s are the components of the correlation matrix $\mathbf{\Sigma}$. Thus, considering elliptical copulas our methods can be extended to the unbalanced longitudinal data where the number of measurements per subject can be different. In the next section we describe the two stage estimation method for our mixture copula based models designed for longitudinal count data.
\section{Parameter estimation}\label{sec5}
The joint probability mass function of $\mathbf{Y}_i$ involves a complex $2n_i$-times folded sum, which is challenging to compute (see, \citet{song2009joint}, \citet{madsen2011joint}). However, for multivariate elliptical copulas, all parameters can be identified from their lower-dimensional sub-copulas, a property that extends to finite mixtures. This can be shown algebraically from the mixture construction. To address computational challenges, composite likelihood methods (CLM) can be used, where pseudolikelihoods are constructed and maximized using consecutive pairs of observations, also known as pairwise likelihood estimation (\citet{le1994logistic}). Recent overviews on CLM are provided in \citet{varin2010mixed} and \citet{varin2011overview}. Unlike full likelihood approaches, CLM only requires specifying bivariate pairs of observations.
\par In order to simplify the computational efforts to estimate the model parameters, we employ the two stage estimation procedure of the composite likelihood as described in \citet{zhao2005composite} and \citet{diao2014composite}. Due to the copula structure of the dependence models, this method produces consistent estimates of the model parameters and their corresponding robust standard errors. Let $f_{ij}(y_{ij}|\theta_{ij})$ be the marginal probability mass function of the response variable $Y_{ij}$ and $\mathbf{\theta}$ be the vector of parameters corresponding to the marginals. Then in the first stage, under working independence assumption we estimate the marginal parameters by maximizing
\begin{equation}\label{marest1}
l(\mathbf{\theta}|\mathbf{y},\mathbf{x}) = \sum_{i=1}^m \sum_{j=1}^{n_i} \log f_{ij}(y_{ij}|\theta_{ij}),  
\end{equation}
and then using the parameter estimates $\hat{\theta}_{ij}$ from (\ref{marest1}), we compute the uniform samples: $u_{ij} = F_{ij}(y_{ij}|\hat{\theta}_{ij})$, $u_{ij}^- = F_{ij}(y_{ij}-1|\hat{\theta}_{ij})$ where $i = 1,\dots,m$, $j = 1,\dots,n_i$. In the second stage, the estimates $\{u_{ij},u_{ij}^-\}$ are inserted into composite likelihood of the form:
\begin{align}\label{copes1}
& l_c(\mathbf{\eta}|\mathbf{u}) = \sum_{i=1}^m \sum_{j=1}^{n_i-1} \sum_{k = j+1}^{n_i} \log P(U_{ij} = u_{ij},U_{ik} = u_{ik}|\eta_{jk}) \nonumber \\ & = \sum_{i=1}^m \sum_{j=1}^{n_i-1} \sum_{k = j+1}^{n_i} \log \Big[C_{\text{mix},2}(u_{ij},u_{ik}|\eta_{jk}) - C_{\text{mix},2}(u_{ij}^-,u_{ik}|\eta_{jk}) - C_{\text{mix},2}(u_{ij},u_{ik}^-|\eta_{jk}) + C_{\text{mix},2}(u_{ij}^-,u_{ik}^-|\eta_{jk})\Big] \nonumber \\ & = \sum_{i=1}^m \sum_{j=1}^{n_i-1} \sum_{k = j+1}^{n_i} \log \Big[\sum_{l=1}^K \pi_l \Big(C_{l,2}(u_{ij},u_{ik}|\phi_{l,jk}) - C_{l,2}(u_{ij}^-,u_{ik}|\phi_{l,jk}) - C_{l,2}(u_{ij},u_{ik}^-|\phi_{l,jk}) + C_{l,2}(u_{ij}^-,u_{ik}^-|\phi_{l,jk})\Big)\Big],
\end{align}
and then maximized with respect to the set of parameters $\mathbf{\eta}$, to obtain the estimates of the association parameters. Here $C_{mix,2}$ and $C_{l,2}$ are the bivariate mixture and component copulas respectively. It has been a standard practice to estimate model parameters of mixture distributions through EM type algorithms. But here we proceed by numerical maximization of the objective functions in (\ref{marest1}) and (\ref{copes1}) with box constraints (see, \citet{macdonald2014numerical}). This method is alternatively known as inference function of margins. The standard errors of the parameter estimates $\hat{\mathbf{\theta}}^* = (\hat{\mathbf{\theta}},\hat{\mathbf{\eta}})^\intercal$ can be numerically obtained from the estimated sandwich information matrix (Godambe information matrix) of the form:
\begin{equation}\label{godmat1}
J(\hat{\mathbf{\theta}}^*) = D(\hat{\mathbf{\theta}}^*)^\intercal M(\hat{\mathbf{\theta}}^*)^{-1} D(\hat{\mathbf{\theta}}^*),
\end{equation}
where $D(\hat{\mathbf{\theta}}^*)$ is a block diagonal matrix and $M(\hat{\mathbf{\theta}}^*)$ is a symmetric positive definite matrix. The explicit forms of these matrices can be found in \citet{zhao2005composite} or \citet{joe2014dependence}. To estimate the parameters we use {\em optim} (\citet{cortez2014modern}) function, and to estimate the information matrix associated with the parameter estimates we use {\em numderiv} (\citet{gilbert2009package}) function in R.
\subsection{Model Validation}\label{sec6}
The validation of copula based regression models has been widely discussed in the literature (see, \citet{genest2009goodness} or \citet{ko2019copula}). To determine which model is best suited for a data set, standard model selection techniques such as AIC (Akaike information criterion) and BIC (Bayesian information criterion) are often used. \citet{varin2005note} and \citet{gao2010composite} modified these two in the case of composite likelihood estimation. These are defined as:
\begin{align}\label{sel1}
CLAIC & = -2l_c(\hat{\mathbf{\theta}}^*) + 2tr(M(\hat{\mathbf{\theta}}^*)D(\hat{\mathbf{\theta}}^*)^{-1}), \nonumber \\ CLBIC & = -2l_c(\hat{\mathbf{\theta}}^*) + \log(m)tr(M(\hat{\mathbf{\theta}}^*)D(\hat{\mathbf{\theta}}^*)^{-1}).
\end{align}
In our applications, model with lowest value of this criteria is selected for drawing conclusions in real data applications. Note that the matrices given in the expression of \ref{sel1} are not same as in \ref{godmat1}. But here we use the exact formulation of these two as described in the previously mentioned papers by plugging in the two stage estimates assuming those are very close to the one step estimates (see, \citet{diao2014composite}). Since we consider mixture of elliptical copulas to model temporal dependencies of count longitudinal data, additionally we can validate the model assumptions utilizing the $t$-plot method by \citet{sun2008heavy} and \citet{shi2012multivariate}.
\par We propose a simple modification of the $t$-plot method for mixture of elliptical copulas utilizing the definition of the mixture distributions. This method was designed to test the elliptical symmetry of multivariate distributions (see \citet{li1997some}), using invariant statistics under orthogonal transformations. \citet{shi2014longitudinal} used this method to validated elliptical assumption of the underlying copula for unbalanced longitudinal count data. The null hypothesis of a $t$-plot is that a sample is from an elliptical multivariate distribution. Such hypothesis could be tested for mixture of elliptical distributions as well. The procedure is given as follows;
\begin{itemize}
    \item For each unit $i$, transform the count variables on uniform scale by $\hat{u}_{ij} = F_{ij}(y_{ij}|\hat{\theta}_{ij})$ for $j = 1,\dots,n_i$. Where $\hat{\theta}_{ij}$ is the estimated marginal parameter from the first stage. Now the transformed data can be considered as a realization of the mixture of elliptical copulas.
    \item Compute the quantiles of $\hat{u}_{ij}$ by $\hat{z}_{ij} = H_j^{-1}(\hat{u}_{ij})$, where $H_j(.)$ denotes the CDF of the $j$-th marginal associated with the elliptical copula. If the copula is well specified, then the vector $\hat{\mathbf{z}}_i = (\hat{z}_{i1},\dots,\hat{z}_{in_i})^\intercal$ would follow a mixture of elliptical distributions.
    \item Let $d_{il}$ be an indicator representing whether the $i$-th unit comes from the $l$-th component of the finite mixture of elliptical distributions as
    \begin{equation}
    F_{\text{mix},n_i} = \sum_{l=1}^K \pi_l F_{l,n_i}(.|\mathbf{\Sigma}_l), \quad \sum_{l=1}^K \pi_l = 1, \pi_l \geq 0, \quad \forall \; l = 1,\dots,K.
    \end{equation}
    and $w_{il}$ be the expected value of $d_{il}$. This can also be interpreted as the posterior probability of $i$-th unit belonging to the $l$-th component of the mixture. Hence we estimate $w_{il}$ by
    \begin{equation}
    \hat{w}_{il} = \frac{\pi_l f_{l,n_i}(\hat{\mathbf{z}}_{i}|\hat{\mathbf{\Sigma}}_l)}{\sum_{l=1}^K \pi_l f_{l,n_i}(\hat{\mathbf{z}}_{i}|\hat{\mathbf{\Sigma}}_l)}, \quad \forall \; l = 1,\dots,K,
    \end{equation}
    Where $\hat{\mathbf{\Sigma}}_l$ is the estimated value of $\mathbf{\Sigma}_l$. Now the $i$-th units is most likely to be coming from the $l$-th component of the mixture if $\hat{w}_{il} = \max\{\hat{w}_{i1},\dots,\hat{w}_{iK}\}$.
    \item After identifying the $l$-th component distribution for the $i$-th unit, we calculate the vector $\hat{\mathbf{z}}_i^* = \hat{\mathbf{\Sigma}}_l^{-1/2}\hat{\mathbf{z}}_i$ and construct the $t$-statistic as
    \begin{equation}
    t_i(\hat{\mathbf{z}}_i^*) = \frac{\sqrt{n_i}\bar{\hat{z}}_i^*}{\sqrt{(n_i - 1)^{-1}\sum_{j=1}^{n_i}(\hat{z}_{ij}^* - \bar{\hat{z}}_i^*)^2}},
    \end{equation}
    where $\bar{\hat{z}}_i^* = n_i^{-1}\sum_{j=1}^{n_i}\hat{z}_{ij}^*$. Thus $t_i(\hat{\mathbf{z}}_i^*)$ should be from a standard $t$-distribution with $n_i - 1$ degrees of freedom.
    \item Repeat the above procedure for all units in the sample and define the transformed variable $v_i = T_1(t_i(\hat{\mathbf{z}}_i^*)|n_i - 1)$ for $i = 1,\dots,m$. If the copula captures the dependence structure properly, then $\mathbf{v} = (v_1,\dots,v_m)^\intercal$ should be a random sample from $U(0,1)$ distribution. Therefore, we plot the sample quantiles of $\mathbf{v}$ against the theoretical quantiles to graphically visualize the goodness-of-fit of the model.
\end{itemize}
\section{Simulation studies}\label{sec7}
In this section we undertake some simulations to monitor the finite sample performance of the proposed mixture copula models designed for count longitudinal data. In our simulations as well as applications following (\citet{arakelian2014clustering}), we set the number of components of the mixture to $K = 2$. We consider two marginal distributions as discussed in Section \ref{sec4}. The specifications of the marginal models are given as
\begin{equation}
\mu_{ij} = \exp(\beta_0 + x_{i1}\beta_1 + x_{i2}\beta_2 + t_{ij}\beta_3), \quad j = 1,\dots,4,
\end{equation}
where the dimension of each units is fixed to $n_i = 4$. We assign same values of the regression coefficients for both of this marginals as, $\beta_0 = 1.0, \beta_1 = 0.5, \beta_2 = 0.5, \beta_3 = -0.5$ and set the over dispersion parameter for Negative Binomial marginals to $\psi = 4.0$. The fixed covariates are generated as $x_{i1} \sim Ber(p = 0.5)$, $x_{i2} \sim dUnif(1,\dots,4)$ (discrete uniform distribution) and the time points $t_{ij} = j$ for $j = 1,\dots,4$, $i = 1,\dots,m$, respectively. We consider two different sample sizes in our simulations as $m = \{200,500\}$. For the mixture copula models we consider one component copula to be autoregressive and another to be exchangeable (parameterized as $\rho = \exp(-\xi)$). We set the parameters for each multivariate copulas as $\xi_1 = 0.3$ (under AR(1) structure) and $\xi_2 = 0.7$ (under EX structure) which are correlation parameters as described in \ref{sec4}, where $\rho_2 = \exp(-\xi_2)$. We also consider $3$ choices for the mixing proportions as $\pi = \{0.25,0.50,0.75\}$. This parameter essentially interprets which correlation structure is present in the data set in what proportion. This $3$-set parameter allows for a wide range of dependence in class of models. Degrees of freedom of the Student-$t$ mixture copulas is kept to $\nu = 4$ to introduce a strong tail dependence. Generating the data sets using standard probability transformations method, we estimate the model parameters using the two stage method discussed in Section \ref{sec5}. We repeat the whole process for $N = 500$ times and report the findings.
\begin{table}[]
    \centering
    \begin{small}
    \rotatebox{90}{
    \begin{minipage}{1.0\textwidth}
    \scalebox{1.0}{
    \tabcolsep = 0.18cm
    \begin{tabular}{|c c|c c c c c|c c c c c|}
    \hline
    \multicolumn{2}{|c|}{} & \multicolumn{5}{c|}{\textbf{m} = 200} & \multicolumn{5}{c|}{\textbf{m} = 500} \\
    \hline
    \textbf{Parameters} & \textbf{True Value} & Mean & Bias & SD & SE & RMSE & Mean & Bias & SD & SE & RMSE \\
    \hline
    $\pi$ & 0.25 & 0.2555 & 0.0055 & 0.1132 & 0.1333 & 0.1133 & 0.2549 & 0.0049 & 0.0759 & 0.0843 & 0.0761 \\
    $\beta_0$ & 1.0 & 0.9964 & -0.0036 & 0.0893 & 0.0856 & 0.0894 & 1.0006 & 0.0006 & 0.0559 & 0.0543 & 0.0559 \\
    $\beta_1$ & 0.5 & 0.5011 & 0.0011 & 0.0525 & 0.0506 & 0.0525 & 0.5008 & 0.0008 & 0.0321 & 0.0323 & 0.0322 \\
    $\beta_2$ & 0.5 & 0.5006 & 0.0006 & 0.0257 & 0.0242 & 0.0257 & 0.4993 & -0.0007 & 0.0160 & 0.0154 & 0.0160 \\
    $\beta_3$ & -0.5 & -0.5004 & -0.0004 & 0.0122 & 0.0132 & 0.0122 & -0.5005 & -0.0005 & 0.0081 & 0.0084 & 0.0081 \\
    $\xi_1$ & 0.3 & 0.3570 & 0.0570 & 0.1687 & 0.1478 & 0.1781 & 0.3302 & 0.0302 & 0.1156 & 0.0958 & 0.1195 \\
    $\xi_2$ & 0.7 & 0.6935 & -0.0065 & 0.1018 & 0.1075 & 0.1020 & 0.6934 & -0.0066 & 0.0755 & 0.0737 & 0.0757 \\
    \hline
    $\pi$ & 0.50 & 0.4884 & -0.0116 & 0.1182 & 0.1289 & 0.1188 & 0.4929 & -0.0071 & 0.0816 & 0.0819 & 0.0819 \\
    $\beta_0$ & 1.0 & 0.9966 & -0.0034 & 0.0879 & 0.0878 & 0.0880 & 1.0016 & 0.0016 & 0.0544 & 0.0554 & 0.0545 \\
    $\beta_1$ & 0.5 & 0.5027 & 0.0027 & 0.0526 & 0.0517 & 0.0527 & 0.5017 & 0.0017 & 0.0342 & 0.0328 & 0.0343 \\
    $\beta_2$ & 0.5 & 0.5003 & 0.0003 & 0.0254 & 0.0249 & 0.0254 & 0.4989 & -0.0011 & 0.0155 & 0.0157 & 0.0156 \\
    $\beta_3$ & -0.5 & -0.5003 & -0.0003 & 0.0135 & 0.0136 & 0.0135 & -0.4996 & 0.0004 & 0.0087 & 0.0086 & 0.0087 \\
    $\xi_1$ & 0.3 & 0.3179 & 0.0179 & 0.0969 & 0.0721 & 0.0985 & 0.3109 & 0.0109 & 0.0593 & 0.0478 & 0.0603 \\
    $\xi_2$ & 0.7 & 0.6862 & -0.0138 & 0.1411 & 0.1404 & 0.1417 & 0.6963 & -0.0037 & 0.1035 & 0.0994 & 0.1036 \\
    \hline
     $\pi$ & 0.75 & 0.7247 & -0.0253 & 0.1119 & 0.1243 & 0.1147 & 0.7502 & 0.0002 & 0.0687 & 0.0772 & 0.0687 \\
    $\beta_0$ & 1.0 & 1.0023 & 0.0023 & 0.0900 & 0.0888 & 0.0901 & 1.0004 & 0.0004 & 0.0586 & 0.0566 & 0.0586 \\
    $\beta_1$ & 0.5 & 0.4985 & -0.0015 & 0.0531 & 0.0527 & 0.0531 & 0.4997 & -0.0003 & 0.0351 & 0.0335 & 0.0352 \\
    $\beta_2$ & 0.5 & 0.4988 & -0.0012 & 0.0253 & 0.0252 & 0.0253 & 0.4998 & -0.0002 & 0.0162 & 0.0160 & 0.0162 \\
    $\beta_3$ & -0.5 & -0.4993 & 0.0007 & 0.0151 & 0.0139 & 0.0151 & -0.5001 & -0.0001 & 0.0090 & 0.0089 & 0.0090 \\
    $\xi_1$ & 0.3 & 0.3085 & 0.0085 & 0.0532 & 0.0502 & 0.0539 & 0.3077 & 0.0077 & 0.0348 & 0.0302 & 0.0356 \\
    $\xi_2$ & 0.7 & 0.6607 & -0.0393 & 0.1775 & 0.2173 & 0.1818 & 0.6844 & -0.0156 & 0.1434 & 0.1654 & 0.1443 \\
    \hline
    \end{tabular}}
    \caption{Parameter estimation using two stage composite likelihood method for Gaussian mixture copula model with Poisson marginals for $N = 500$ simulated data sets with two different sample sizes.}
    \label{tab:simulation1}
    \end{minipage}}
    \end{small}
\end{table}
\begin{table}[]
    \centering
    \begin{small}
    \rotatebox{90}{
    \begin{minipage}{1.0\textwidth}
    \scalebox{1.0}{
    \tabcolsep = 0.18cm
    \begin{tabular}{|c c|c c c c c|c c c c c|}
    \hline
    \multicolumn{2}{|c|}{} & \multicolumn{5}{c|}{\textbf{m} = 200} & \multicolumn{5}{c|}{\textbf{m} = 500} \\
    \hline
    \textbf{Parameters} & \textbf{True Value} & Mean & Bias & SD & SE & RMSE & Mean & Bias & SD & SE & RMSE \\
    \hline
    $\pi$ & 0.25 & 0.2695 & 0.0195 & 0.1233 & 0.1545 & 0.1249 & 0.2495 & -0.0005 & 0.0865 & 0.0960 & 0.0865 \\
    $\beta_0$ & 1.0 & 0.9958 & -0.0042 & 0.0859 & 0.0857 & 0.0860 & 0.9995 & -0.0005 & 0.0536 & 0.0541 & 0.0536 \\
    $\beta_1$ & 0.5 & 0.4981 & -0.0019 & 0.0518 & 0.0504 & 0.0518 & 0.5040 & 0.0040 & 0.0319 & 0.0320 & 0.0321 \\
    $\beta_2$ & 0.5 & 0.5018 & 0.0018 & 0.0236 & 0.0242 & 0.0237 & 0.4990 & -0.0010 & 0.0150 & 0.0153 & 0.0150 \\
    $\beta_3$ & -0.5 & -0.5004 & -0.0004 & 0.0127 & 0.0132 & 0.0127 & -0.4997 & 0.0003 & 0.0081 & 0.0084 & 0.0081 \\
    $\xi_1$ & 0.3 & 0.3525 & 0.0525 & 0.1959 & 0.2016 & 0.2028 & 0.3232 & 0.0232 & 0.1353 & 0.1457 & 0.1373 \\
    $\xi_2$ & 0.7 & 0.7006 & 0.0006 & 0.1291 & 0.1580 & 0.1291 & 0.7047 & 0.0047 & 0.1044 & 0.1166 & 0.1045 \\
    \hline
    $\pi$ & 0.50 & 0.5003 & 0.0003 & 0.1329 & 0.1479 & 0.1329 & 0.4937 & -0.0063 & 0.0886 & 0.0920 & 0.0888 \\
    $\beta_0$ & 1.0 & 0.9998 & -0.0002 & 0.0858 & 0.0812 & 0.0858 & 1.0019 & 0.0019 & 0.0578 & 0.0554 & 0.0578 \\
    $\beta_1$ & 0.5 & 0.5020 & 0.0020 & 0.0516 & 0.0516 & 0.0517 & 0.4968 & -0.0032 & 0.0338 & 0.0328 & 0.0340 \\
    $\beta_2$ & 0.5 & 0.4992 & -0.0008 & 0.0240 & 0.0245 & 0.0240 & 0.4998 & -0.0002 & 0.0159 & 0.0157 & 0.0159 \\
    $\beta_3$ & -0.5 & -0.4994 & 0.0006 & 0.0137 & 0.0136 & 0.0137 & -0.4996 & 0.0004 & 0.0082 & 0.0086 & 0.0082 \\
    $\xi_1$ & 0.3 & 0.3237 & 0.0237 & 0.1249 & 0.1144 & 0.1271 & 0.3055 & 0.0055 & 0.0682 & 0.0699 & 0.0684 \\
    $\xi_2$ & 0.7 & 0.6899 & -0.0101 & 0.1666 & 0.2288 & 0.1669 & 0.7042 & 0.0042 & 0.1245 & 0.1580 & 0.1246 \\
    \hline
     $\pi$ & 0.75 & 0.7222 & -0.0278 & 0.1223 & 0.1546 & 0.1255 & 0.7320 & -0.0180 & 0.0841 & 0.0928 & 0.0860 \\
    $\beta_0$ & 1.0 & 0.9967 & -0.0033 & 0.0942 & 0.0889 & 0.0943 & 1.0041 & 0.0041 & 0.0575 & 0.0563 & 0.0576 \\
    $\beta_1$ & 0.5 & 0.5033 & 0.0033 & 0.0521 & 0.0524 & 0.0522 & 0.4989 & -0.0011 & 0.0318 & 0.0333 & 0.0318 \\
    $\beta_2$ & 0.5 & 0.5006 & 0.0006 & 0.0260 & 0.0250 & 0.0260 & 0.4988 & -0.0012 & 0.0162 & 0.0159 & 0.0162 \\
    $\beta_3$ & -0.5 & -0.5010 & -0.0010 & 0.0149 & 0.0140 & 0.0149 & -0.4996 & 0.0004 & 0.0089 & 0.0089 & 0.0089 \\
    $\xi_1$ & 0.3 & 0.3119 & 0.0119 & 0.0627 & 0.0891 & 0.0638 & 0.3057 & 0.0057 & 0.0434 & 0.0512 & 0.0438 \\
    $\xi_2$ & 0.7 & 0.6617 & -0.0383 & 0.1894 & 0.2880 & 0.1932 & 0.6714 & -0.0286 & 0.1535 & 0.2145 & 0.1561 \\
    \hline
    \end{tabular}}
    \caption{Parameter estimation using two stage composite likelihood method for Student-$t$ ($\nu = 4$) mixture copula model with Poisson marginals for $N = 500$ simulated data sets with two different sample sizes.}
    \label{tab:simulation2}
    \end{minipage}}
    \end{small}
\end{table}
\begin{table}[]
    \centering
    \begin{small}
    \rotatebox{90}{
    \begin{minipage}{1.0\textwidth}
    \scalebox{1.0}{
    \tabcolsep = 0.18cm
    \begin{tabular}{|c c|c c c c c|c c c c c|}
    \hline
    \multicolumn{2}{|c|}{} & \multicolumn{5}{c|}{\textbf{m} = 200} & \multicolumn{5}{c|}{\textbf{m} = 500} \\
    \hline
    \textbf{Parameters} & \textbf{True Value} & Mean & Bias & SD & SE & RMSE & Mean & Bias & SD & SE & RMSE \\
    \hline
    $\pi$ & 0.25 & 0.2611 & 0.0111 & 0.1188 & 0.1314 & 0.1193 & 0.2466 & -0.0034 & 0.0775 & 0.0832 & 0.0775 \\
    $\beta_0$ & 1.0 & 1.0117 & 0.0117 & 0.1153 & 0.1212 & 0.1159 & 0.9975 & -0.0025 & 0.0771 & 0.0774 & 0.0771 \\
    $\beta_1$ & 0.5 & 0.4933 & -0.0067 & 0.0800 & 0.0808 & 0.0802 & 0.5034 & 0.0034 & 0.0517 & 0.0514 & 0.0518 \\
    $\beta_2$ & 0.5 & 0.4985 & -0.0015 & 0.0366 & 0.0370 & 0.0366 & 0.4992 & -0.0008 & 0.0230 & 0.0236 & 0.0230 \\
    $\beta_3$ & -0.5 & -0.5017 & -0.0017 & 0.0189 & 0.0186 & 0.0190 & -0.4997 & 0.0003 & 0.0125 & 0.0117 & 0.0125 \\
    $\psi$ & 4.0 & 4.1369 & 0.1369 & 0.5739 & 0.5530 & 0.5900 & 4.0512 & 0.0512 & 0.3504 & 0.3434 & 0.3541 \\
    $\xi_1$ & 0.3 & 0.3555 & 0.0555 & 0.1738 & 0.1348 & 0.1825 & 0.3247 & 0.0247 & 0.1217 & 0.0953 & 0.1242 \\
    $\xi_2$ & 0.7 & 0.6994 & -0.0006 & 0.1124 & 0.1088 & 0.1124 & 0.7006 & 0.0006 & 0.0785 & 0.0721 & 0.0785 \\
    \hline
    $\pi$ & 0.50 & 0.4910 & -0.0090 & 0.1191 & 0.1297 & 0.1195 & 0.4926 & -0.0074 & 0.0753 & 0.0813 & 0.0757 \\
    $\beta_0$ & 1.0 & 1.0040 & 0.0040 & 0.1243 & 0.1236 & 0.1243 & 0.9998 & -0.0002 & 0.0788 & 0.0789 & 0.0788 \\
    $\beta_1$ & 0.5 & 0.4912 & -0.0088 & 0.0824 & 0.0822 & 0.0829 & 0.5015 & 0.0015 & 0.0529 & 0.0524 & 0.0529 \\
    $\beta_2$ & 0.5 & 0.5001 & 0.0001 & 0.0370 & 0.0376 & 0.0370 & 0.4995 & -0.0005 & 0.0237 & 0.0240 & 0.0237 \\
    $\beta_3$ & -0.5 & -0.5014 & -0.0014 & 0.0194 & 0.0192 & 0.0195 & -0.4995 & 0.0005 & 0.0127 & 0.0122 & 0.0127 \\
    $\psi$ & 4.0 & 4.2174 & 0.2174 & 0.6299 & 0.5866 & 0.6663 & 4.0615 & 0.0615 & 0.3463 & 0.3565 & 0.3518 \\
    $\xi_1$ & 0.3 & 0.3315 & 0.0315 & 0.1139 & 0.0783 & 0.1181 & 0.3087 & 0.0087 & 0.0617 & 0.0488 & 0.0623 \\
    $\xi_2$ & 0.7 & 0.6858 & -0.0142 & 0.1448 & 0.1460 & 0.1455 & 0.7008 & 0.0008 & 0.1054 & 0.1008 & 0.1054 \\
    \hline
     $\pi$ & 0.75 & 0.7221 & -0.0279 & 0.1174 & 0.1228 & 0.1207 & 0.7422 & -0.0078 & 0.0745 & 0.0777 & 0.0749 \\
    $\beta_0$ & 1.0 & 1.0007 & 0.0007 & 0.1322 & 0.1267 & 0.1322 & 0.9969 & -0.0031 & 0.0827 & 0.0803 & 0.0828 \\
    $\beta_1$ & 0.5 & 0.4999 & -0.0001 & 0.0832 & 0.0840 & 0.0832 & 0.5005 & 0.0005 & 0.0507 & 0.0533 & 0.0507 \\
    $\beta_2$ & 0.5 & 0.4980 & -0.0020 & 0.0409 & 0.0383 & 0.0409 & 0.5002 & 0.0002 & 0.0260 & 0.0245 & 0.0260 \\
    $\beta_3$ & -0.5 & -0.5010 & -0.0010 & 0.0191 & 0.0198 & 0.0191 & -0.4995 & 0.0005 & 0.0124 & 0.0125 & 0.0124 \\
    $\psi$ & 4.0 & 4.1623 & 0.1623 & 0.6292 & 0.5924 & 0.6498 & 4.0639 & 0.0639 & 0.3554 & 0.3677 & 0.3611 \\
    $\xi_1$ & 0.3 & 0.3207 & 0.0207 & 0.0616 & 0.0528 & 0.0650 & 0.3080 & 0.0080 & 0.0401 & 0.0308 & 0.0409 \\
    $\xi_2$ & 0.7 & 0.6518 & -0.0482 & 0.1770 & 0.2146 & 0.1834 & 0.6816 & -0.0184 & 0.1456 & 0.1524 & 0.1467 \\
    \hline
    \end{tabular}}
    \caption{Parameter estimation using two stage composite likelihood method for Gaussian mixture copula model with Negative Binomial marginals for $N = 500$ simulated data sets with two different sample sizes.}
    \label{tab:simulation3}
    \end{minipage}}
    \end{small}
\end{table}
\begin{table}[]
    \centering
    \begin{small}
    \rotatebox{90}{
    \begin{minipage}{1.0\textwidth}
    \scalebox{1.0}{
    \tabcolsep = 0.18cm
    \begin{tabular}{|c c|c c c c c|c c c c c|}
    \hline
    \multicolumn{2}{|c|}{} & \multicolumn{5}{c|}{\textbf{m} = 200} & \multicolumn{5}{c|}{\textbf{m} = 500} \\
    \hline
    \textbf{Parameters} & \textbf{True Value} & Mean & Bias & SD & SE & RMSE & Mean & Bias & SD & SE & RMSE \\
    \hline
    $\pi$ & 0.25 & 0.2651 & 0.0151 & 0.1189 & 0.1413 & 0.1199 & 0.2500 & 0.0000 & 0.0823 & 0.0947 & 0.0823 \\
    $\beta_0$ & 1.0 & 0.9921 & -0.0079 & 0.1229 & 0.1221 & 0.1231 & 0.9956 & -0.0044 & 0.0812 & 0.0776 & 0.0814 \\
    $\beta_1$ & 0.5 & 0.4983 & -0.0017 & 0.0851 & 0.0811 & 0.0851 & 0.5018 & 0.0018 & 0.0540 & 0.0515 & 0.0541 \\
    $\beta_2$ & 0.5 & 0.5012 & 0.0012 & 0.0385 & 0.0372 & 0.0385 & 0.5006 & 0.0006 & 0.0244 & 0.0237 & 0.0244 \\
    $\beta_3$ & -0.5 & -0.4991 & 0.0009 & 0.0179 & 0.0184 & 0.0180 & -0.5000 & 0.0000 & 0.0121 & 0.0117 & 0.0121 \\
    $\psi$ & 4.0 & 4.1597 & 0.1597 & 0.6878 & 0.6176 & 0.7061 & 4.0586 & 0.0586 & 0.3596 & 0.3803 & 0.3643 \\
    $\xi_1$ & 0.3 & 0.3580 & 0.0580 & 0.1988 & 0.2019 & 0.2071 & 0.3168 & 0.0168 & 0.1362 & 0.1314 & 0.1372 \\
    $\xi_2$ & 0.7 & 0.7010 & 0.0010 & 0.1237 & 0.1486 & 0.1237 & 0.7053 & 0.0053 & 0.0947 & 0.1018 & 0.0949 \\
    \hline
    $\pi$ & 0.50 & 0.5005 & 0.0005 & 0.1375 & 0.1444 & 0.1375 & 0.4974 & -0.0026 & 0.0906 & 0.0909 & 0.0906 \\
    $\beta_0$ & 1.0 & 0.9911 & -0.0089 & 0.1305 & 0.1242 & 0.1308 & 1.0069 & 0.0069 & 0.0811 & 0.0790 & 0.0814 \\
    $\beta_1$ & 0.5 & 0.5044 & 0.0044 & 0.0836 & 0.0824 & 0.0837 & 0.4966 & -0.0034 & 0.0509 & 0.0523 & 0.0510 \\
    $\beta_2$ & 0.5 & 0.5000 & 0.0000 & 0.0386 & 0.0377 & 0.0386 & 0.4988 & -0.0012 & 0.0244 & 0.0241 & 0.0244 \\
    $\beta_3$ & -0.5 & -0.4996 & 0.0004 & 0.0184 & 0.0191 & 0.0184 & -0.5007 & -0.0007 & 0.0119 & 0.0121 & 0.0119 \\
    $\psi$ & 4.0 & 4.1899 & 0.1899 & 0.6372 & 0.6316 & 0.6649 & 4.0948 & 0.0948 & 0.3862 & 0.3930 & 0.3977 \\
    $\xi_1$ & 0.3 & 0.3313 & 0.0313 & 0.1208 & 0.1068 & 0.1247 & 0.3120 & 0.0120 & 0.0756 & 0.0692 & 0.0765 \\
    $\xi_2$ & 0.7 & 0.6936 & -0.0064 & 0.1579 & 0.2037 & 0.1580 & 0.7022 & 0.0022 & 0.1310 & 0.1519 & 0.1310 \\
    \hline
     $\pi$ & 0.75 & 0.7192 & -0.0308 & 0.1232 & 0.1462 & 0.1270 & 0.7490 & -0.0010 & 0.0748 & 0.0950 & 0.0748 \\
    $\beta_0$ & 1.0 & 0.9916 & -0.0084 & 0.1266 & 0.1258 & 0.1266 & 0.9999 & -0.0001 & 0.0782 & 0.0802 & 0.0782 \\
    $\beta_1$ & 0.5 & 0.5005 & 0.0005 & 0.0793 & 0.0836 & 0.0793 & 0.5002 & 0.0002 & 0.0534 & 0.0534 & 0.0534 \\
    $\beta_2$ & 0.5 & 0.5023 & 0.0023 & 0.0392 & 0.0384 & 0.0392 & 0.5005 & 0.0005 & 0.0241 & 0.0244 & 0.0241 \\
    $\beta_3$ & -0.5 & -0.4999 & 0.0001 & 0.0202 & 0.0196 & 0.0202 & -0.5005 & -0.0005 & 0.0123 & 0.0125 & 0.0123 \\
    $\psi$ & 4.0 & 4.2224 & 0.2224 & 0.6987 & 0.6535 & 0.7332 & 4.0906 & 0.0906 & 0.4264 & 0.4002 & 0.4359 \\
    $\xi_1$ & 0.3 & 0.3180 & 0.0180 & 0.0711 & 0.0824 & 0.0733 & 0.3084 & 0.0084 & 0.0437 & 0.0540 & 0.0445 \\
    $\xi_2$ & 0.7 & 0.6743 & -0.0257 & 0.1780 & 0.2278 & 0.1799 & 0.6863 & -0.0137 & 0.1464 & 0.1606 & 0.1470 \\
    \hline
    \end{tabular}}
    \caption{Parameter estimation using two stage composite likelihood method for Student-$t$ ($\nu = 4$) mixture copula model with Negative Binomial marginals for $N = 500$ simulated data sets with two different sample sizes.}
    \label{tab:simulation4}
    \end{minipage}}
    \end{small}
\end{table}
\par Table \ref{tab:simulation1}, \ref{tab:simulation2}, \ref{tab:simulation3} and \ref{tab:simulation4} presents the simulation results for the considered models. Within each table, we report the mean, the biases [$\frac{1}{N}\sum_{i=1}^N (\hat{\mathbf{\theta}}_j^* - \mathbf{\theta}^*)$], empirical standard deviations (denoted as SD), average standard errors obtained from the asymptotic covariance matrices (denoted as SE) and roots of mean square errors [$\sqrt{\frac{1}{N}\sum_{i=1}^N (\hat{\mathbf{\theta}}_j^* - \mathbf{\theta}^*)^2}$], where $\hat{\mathbf{\theta}}_j^*$ is the parameter estimates for the $j$-th sample. The average estimates are very close to the corresponding true parameters for both sample sizes. The results show consistent performance of the proposed models with two stage composite likelihood estimation as the biases and roots of the mean square errors decrease with increasing sample size. The average of the asymptotic standard errors of the parameter estimates (SE) is comparable with the empirical standard deviation (SD) of point estimates, indicating the accuracy of the uncertainty estimates. The standard errors of the regression parameter estimates are larger in the Negative Binomial based models than of the Poisson based models. We see that as the mixing proportion increases for a mixture component the corresponding bias and RMSE decreases. Student-$t$ copula implies a stronger dependence than Gaussian copula which inflicted in increased bias and uncertainty of the estimates for a given sample size. Overall the simulation results suggest that the estimation method for the proposed class of models provides a valid basis for inference. In the next section we apply these models to analyze two real world data sets.
\section{Applications}\label{sec8}
We illustrate the effectiveness of our mixture copula based count regression models onto some real world data sets. These data sets are publicly available in \citet{sutradhar2011dynamic} and \citet{thall1990some}.
\subsection{The health care utilization data}\label{subsec81}
We consider the data set on the health care utilization problem collected by the General Hospital of the city of St. John's, Newfoundland, Canada. The data refer to the number of physician visits over the years from $1985$ to $1990$ for $180$ individuals (as given in Appendix $6A$ of \citet{sutradhar2011dynamic}). The information on $4$ covariates, namely, gender, number of chronic conditions, educational level and age were recorded for each individual. It is appropriate to assume that the repeated counts recorded for $6$ years will be longitudinally correlated. We are interested in estimating the temporal dependency of the count responses after taking the effects of covariates into account. The summary statistics for these responses are presented in \ref{tab:summary1}, which indicates that the data is not significantly overdispersed. Different correlation structures such as EX or AR(1) are considered in \citet{sutradhar2011dynamic} and the model parameters were estimated using generalized quasi likelihood (GQL) method. However in our mixture copula based modeling framework we use AR(1) and EX structure in each copula components. Following the notations used in Section \ref{sec4}, we consider the covariates as sex ($0 =$ male, $1 =$ female), chronic disease status (number of active diseases from $0$ to $4$), education level ($1$ for less than high school, $2$ for high school, $3$ for university graduate and $4$ for post graduate) and age of an individual (with deviation of $50$ years). We consider the mean function as
\begin{equation}\label{dreal1}
\mu_{ij} = \exp(\beta_0 + \text{sex}_i\beta_1 + \text{crn}_i\beta_2 + \text{edu}_i\beta_3 + \text{age}_i\beta_4 + t_{ij}\beta_5), \quad j = 1,\dots,6, 
\end{equation}
where $t_{ij}$ is the respective year of visit from $1$ to $6$. From the empirical correlation matrix of the count measurements, EX structure is seems to be appropriate for this data set. We apply our methods discussed in Section \ref{sec5} to estimate the regression as well as dependence parameters with $2$ choices for the marginal distributions, standard and $2$-component mixture of elliptical copulas.
\begin{table}[h]
    \centering
    \begin{small}
    \scalebox{1.0}{
    \tabcolsep = 0.18cm
    \begin{tabular}{|c|c c c c c c|}
    \hline
    & Visit 1 & Visit 2 & Visit 3 & Visit 4 & Visit 5 & Visit 6 \\
    \hline
    Mean & 3.9333 & 3.8833 & 3.9667 & 4.7000 & 5.4111 & 4.7556 \\
    SD & 4.6756 & 4.4159 & 4.8079 & 5.5922 & 6.2327 & 5.5266 \\
    \hline
    \end{tabular}}
    \caption{Summary statistics for $6$ count responses across time for $180$ subjects.}
    \label{tab:summary1}
    \end{small}
\end{table}
\begin{table}[h]
    \centering
    \begin{small}
    \scalebox{1.0}{
    \tabcolsep = 0.18cm
    \begin{tabular}{|c|c c|c|c c|}
    \hline
    \multicolumn{3}{|c}{Poisson marginals} & \multicolumn{3}{|c|}{Negative Binomial marginals} \\
    \hline
    \textbf{Parameters} & Est. & SE & \textbf{Parameters} & Est. & SE \\
    \hline
    $\beta_0$ & 0.7028 & 0.1830 & $\beta_0$ & 0.6727 & 0.1903 \\
    $\beta_1$ & 0.6179 & 0.1379 & $\beta_1$ & 0.6960 & 0.1225 \\
    $\beta_2$ & 0.2003 & 0.0480 & $\beta_2$ & 0.2106 & 0.0497 \\
    $\beta_3$ & 0.0306 & 0.0649 & $\beta_3$ & 0.0280 & 0.0632 \\
    $\beta_4$ & 0.0076 & 0.0042 & $\beta_4$ & 0.0100 & 0.0025 \\
    $\beta_5$ & 0.0608 & 0.0165 & $\beta_5$ & 0.0619 & 0.0176 \\
    - & - & - & $\psi$ & 0.9747 & 0.0954 \\
    \hline
    \end{tabular}}
    \caption{Estimated marginal parameters and their standard errors for the healthcare utilization data, obtained using the model in (\ref{dreal1}) with Poisson and Negative Binomial marginals.}
    \label{tab:realdata1fit1}
    \end{small}
\end{table}
\begin{table}[h]
    \centering
    \begin{small}
    \scalebox{1.0}{
    \tabcolsep=0.18cm
    \begin{tabular}{|c|c|c|c c|c|c|c|}
    \hline
    \textbf{Model} & \textbf{Copula} & \textbf{Parameters} & Est. & SE & Comp-like & CLAIC & CLBIC \\
    \hline
    Poisson & Gaussian & $\xi_2$ & 1.0222 & 0.0532 & -17144.81 & 34486.16 & 34808.81 \\
    & exchangeable & & & & & & \\
    & Gaussian & $\pi$ & 0.5339 & 0.0386 & -16817.61 & 33839.47 & 34165.56 \\
    & mixture & $\xi_1$ & 2.1041 & 0.3020 & & & \\
    & & $\xi_2$ & 0.2348 & 0.0259 & & & \\
    & Student-$t$ ($\nu = 15$) & $\xi_2$ & 1.0291 & 0.0832 & -16932.91 & 34063.10 & 34386.36 \\
    & exchangeable & & & & & & \\
    & Student-$t$ ($\nu = 17$) & $\pi$ & 0.5634 & 0.0461 & \textbf{-16803.36} & \textbf{33810.99} & \textbf{34137.09} \\
    & mixture & $\xi_1$ & 1.6801 & 0.2099 & & & \\
    & & $\xi_2$ & 0.2653 & 0.0419 & & & \\
    \hline
    Negative & Gaussian & $\xi_2$ & 0.6181 & 0.0246 & -13010.82 & 26073.76 & 26158.89 \\
    Binomial & exchangeable & & & & & & \\
    & Gaussian & $\pi$ & 0.4824 & 0.0968 & -12980.33 & 26014.73 & 26101.06 \\
    & mixture & $\xi_1$ & 0.6448 & 0.1003 & & & \\
    & & $\xi_2$ & 0.2780 & 0.0763 & & & \\
    & Student-$t$ ($\mathbf{\nu} = 10$) & $\xi_2$ & 0.6002 & 0.0255 & -12999.87 & 26061.49 & 26130.99 \\
    & exchangeable & & & & & & \\
    & Student-$t$ ($\mathbf{\nu} = 12$) & $\pi$ & 0.4181 & 0.0953 & \textbf{-12973.71} & \textbf{26010.99} & \textbf{26080.55} \\
    & mixture & $\xi_1$ & 0.3137 & 0.1472 & & & \\
    & & $\xi_2$ & 0.5672 & 0.1812 & & & \\
    \hline
    \end{tabular}}
    \end{small}
    \caption{Estimated dependence parameters and their standard errors for the healthcare utilization data, using both standard and mixture elliptical copulas. The maximum composite log-likelihood value, CLAIC, and CLBIC for each model are also reported.}
    \label{tab:realdata1fit2}
\end{table}
\par
\begin{figure}
    \centering
    \includegraphics[width = 12cm]{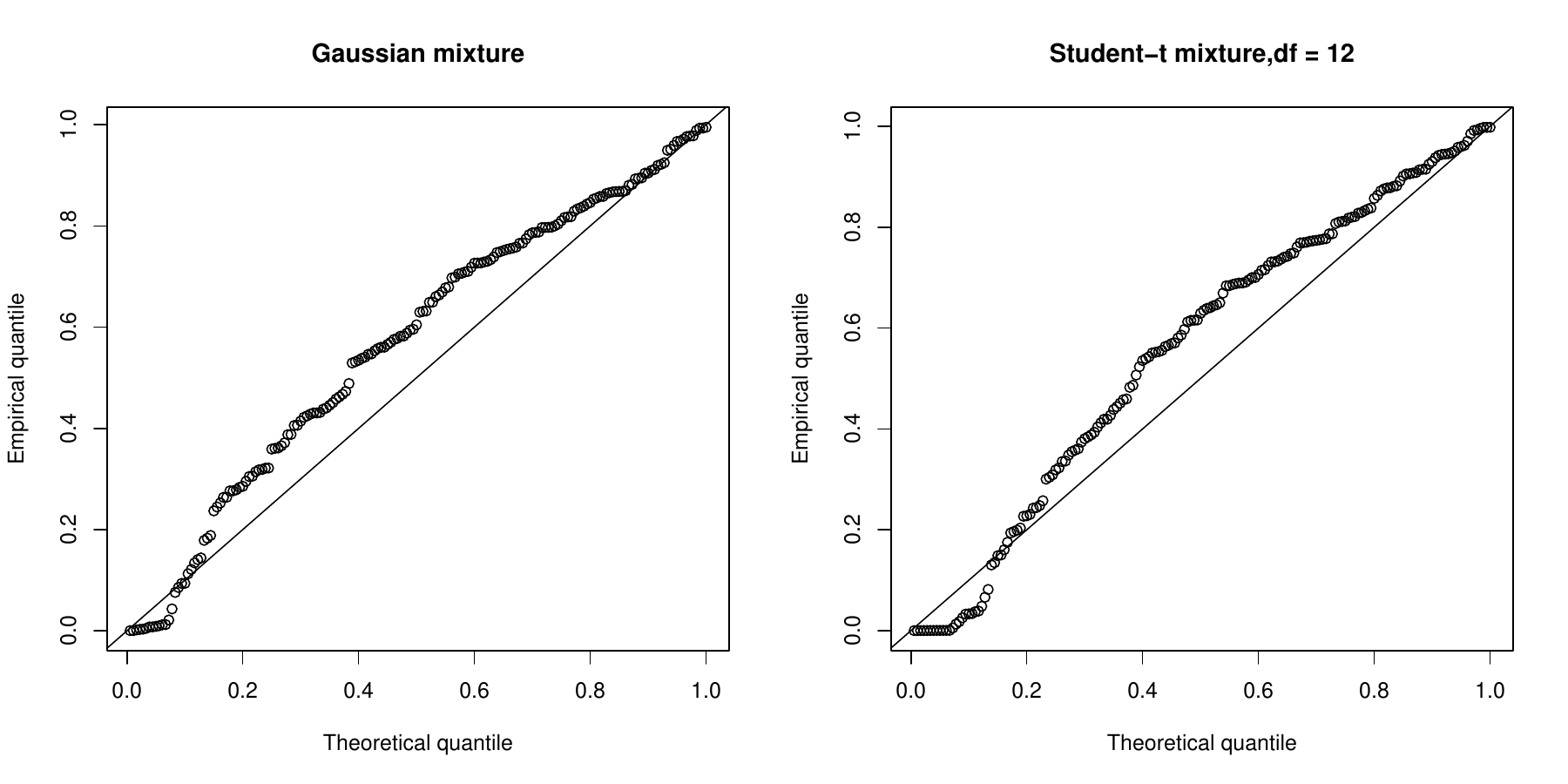}
    \caption{$t$-plots for the mixture of elliptical copulas applied to the healthcare utilization data.}
    \label{fig:data1.t-plot}
\end{figure}

 The results are reported in Table \ref{tab:realdata1fit1} and \ref{tab:realdata1fit2}. Based on the composite log-likelihood values and the selection criteria, the Negative Binomial based model with Student-$t$ mixture copula provide with the best fit. Based on the value of the degrees of freedom parameter $\nu$, moderate level of tail dependence is detected in the data set. Therefore we chose this model to interpret the parameter estimates. The positive value of $\beta_1$ suggests that the females made more visits to the physician as compared to the males. The positive values of $\beta_2$ and $\beta_4$ suggest that individuals having one or more chronic diseases or individuals belonging to the older age group pay more visits to the physicians, as expected. Based on the estimate of $\beta_3$ and its associated standard error, it shows education level has a minor impact on the number of physician visits by the individuals. Since we used the data set for all $6$ available time points, this conclusion differs from that in \citet{sutradhar2003overview}. Finally based on positive value of $\beta_5$, individuals payed more visits to the physician in the subsequent years. Based on the estimates of the dependence parameters from the mixture copula, the value of $\pi$ suggests EX structure is more prominent in the data set, but if we consider the Poisson based model which is seen to be inferior to the Negative Binomial based model, it shows AR(1) structure is more prominent. We utilize the theoretical expressions of Kendall's tau and Spearman's rho derived in Section \ref{sec3}, to obtain the concordance matrices and to compare them with their empirical versions. We calculate the sample Kendall's tau and Spearman's rho values for the residuals of the fitted marginal models. Let $A(\tau)$ and $A(\rho)$ denote the empirical Kendall's tau and Spearman's rho matrices for this data set. Using Theorem \ref{thm3} and Theorem \ref{thm4} we analytically obtain $A(\hat{\tau})$ and $A(\hat{\rho})$ with the estimated parameters. These are given as
\begin{align}\label{matcomp1}
A(\tau) & = \left[\arraycolsep=3.0pt\def\arraystretch{1.0}\begin{array}{cccccc} 1.00 \\ 0.44 & 1.00 \\ 0.41 & 0.40 & 1.00 \\ 0.31 & 0.36 & 0.43 & 1.00 \\ 0.30 & 0.37 & 0.39 & 0.50 & 1.00 \\ 0.27 & 0.27 & 0.33 & 0.43 & 0.48 & 1.00 \end{array}\right], A(\hat{\tau}) = \left[\arraycolsep=3.0pt\def\arraystretch{1.0}\begin{array}{cccccc} 1.00 \\ 0.41 & 1.00 \\ 0.35 & 0.42 & 1.00 \\ 0.31 & 0.35 & 0.42 & 1.00 \\ 0.28 & 0.31 & 0.36 & 0.42 & 1.00 \\ 0.26 & 0.28 & 0.31 & 0.36 & 0.42 & 1.00 \end{array}\right], \nonumber \\ A(\rho) & = \left[\arraycolsep=3.0pt\def\arraystretch{1.0}\begin{array}{cccccc} 1.00 \\ 0.60 & 1.00 \\ 0.56 & 0.55 & 1.00 \\ 0.43 & 0.50 & 0.59 & 1.00 \\ 0.42 & 0.52 & 0.53 & 0.67 & 1.00 \\ 0.38 & 0.40 & 0.46 & 0.60 & 0.60 & 1.00 \end{array}\right] \text{and} \; A(\hat{\rho}) = \left[\arraycolsep=3.0pt\def\arraystretch{1.0}\begin{array}{cccccc} 1.00 \\ 0.59 & 1.00 \\ 0.51 & 0.59 & 1.00 \\ 0.45 & 0.51 & 0.59 & 1.00 \\ 0.41 & 0.45 & 0.51 & 0.59 & 1.00 \\ 0.38 & 0.41 & 0.45 & 0.51 & 0.60 & 1.00 \end{array}\right]. \nonumber
\end{align}
These estimated concordance matrices are very close to their empirical versions and they indicate high longitudinal correlations. To evaluate the goodness-of-fit we implement the modified $t$-plot discussed in Section \ref{sec6} to the mixture copula models with Negative Binomial marginals in Figure \ref{fig:data1.t-plot}. We can see linear trend along the $45$-degree line for both the plots. But for the Student-$t$ mixture copula the uniform quantiles are relatively closer to the line than the Gaussian copula. This suggest that Student-$t$ mixture copula is more suitable for the temporal dependency for this data set.
\subsection{The epilepsy data}
We also consider the well-known longitudinal study on epileptic seizures previously analyzed by several authors such as \citet{thall1990some}, \citet{jowaheer2002analysing} and \citet{masarotto2012gaussian} among others. This study was conducted on $59$ patients suffering from simple or complex partial seizures. Among these patients, $31$ of them were given the anti-epileptic drug progabide and remaining $28$ patients were given placebo randomly. The patients were observed for four successive clinical visits after randomization, and the number of seizures occurred over the previous $2$ weeks for each individual was reported. The summary statistics are presented on Table \ref{tab:summary2}, which indicates high overdispersion in the data. Following \citet{ahmmed2021generalized}, we consider $3$ covariates to analyze this data set as logarithm of baseline seizure count, logarithm of age in years and the treatment indicator ($0 =$ placebo, $1 =$ progabide). From preliminary descriptive analysis, it is evident that the data is over dispersed. We consider the mean function as
\begin{equation}\label{dreal2}
\mu_{ij} = \exp(\beta_0 + \text{lbase}_i\beta_1 + \text{trt}_i\beta_2 + \text{lage}_i\beta_3 + t_{ij}\beta_5), \quad j = 1,\dots,4, 
\end{equation}
where $t_{ij}$ is the respective visit number from $1$ to $6$. For this data set also EX structure seems to be appropriate based on the empirical correlation matrix. We analyze this data set using previously described methodology as well.
\begin{table}[h]
    \centering
    \begin{small}
    \scalebox{1.0}{
    \tabcolsep = 0.18cm
    \begin{tabular}{|c|c c c c|}
    \hline
    & Visit 1 & Visit 2 & Visit 3 & Visit 4 \\
    \hline
    Mean & 8.9491 & 8.3559 & 8.4407 & 7.3390 \\
    SD & 14.8352 & 10.187 & 14.1486 & 9.6356 \\
    \hline
    \end{tabular}}
    \caption{Summary statistics for $4$ count responses across time for $59$ subjects.}
    \label{tab:summary2}
    \end{small}
\end{table}
\begin{table}[h]
    \centering
    \begin{small}
    \scalebox{1.0}{
    \tabcolsep = 0.18cm
    \begin{tabular}{|c|c c|c|c c|}
    \hline
    \multicolumn{3}{|c}{Poisson marginals} & \multicolumn{3}{|c|}{Negative Binomial marginals} \\
    \hline
    \textbf{Parameters} & Est. & SE & \textbf{Parameters} & Est. & SE \\
    \hline
    $\beta_0$ & -3.7090 & 0.9579 & $\beta_0$ & -2.3136 & 0.8877 \\
    $\beta_1$ & 1.2199 & 0.1543 & $\beta_1$ & 1.0536 & 0.1084 \\
    $\beta_2$ & -0.0397 & 0.1869 & $\beta_2$ & -0.2421 & 0.1545 \\
    $\beta_3$ & 0.5184 & 0.2377 & $\beta_3$ & 0.3034 & 0.2312 \\
    $\beta_4$ & -0.0574 & 0.0350 & $\beta_4$ & -0.0542 & 0.0360 \\
    - & - & - & $\psi$ & 2.6101 & 0.5975 \\
    \hline
    \end{tabular}}
    \caption{Estimated marginal parameters and their standard errors for the epilepsy data, obtained using the model in (\ref{dreal2}) with Poisson and Negative Binomial marginals.}
    \label{tab:realdata2fit1}
    \end{small}
\end{table}
\begin{table}[h]
    \centering
    \begin{small}
    \scalebox{1.0}{
    \tabcolsep=0.18cm
    \begin{tabular}{|c|c|c|c c|c|c|c|}
    \hline
    \textbf{Model} & \textbf{Copula} & \textbf{Parameters} & Est. & SE & Comp-like & CLAIC & CLBIC \\
    \hline
    Poisson & Gaussian & $\xi_2$ & 1.1832 & 0.0875 & -2245.24 & 4636.15 & 4789.72 \\
    & exchangeable & & & & & & \\
    & Gaussian & $\pi$ & 0.5603 & 0.0899 & -2221.13 & 4590.23 & 4743.93 \\
    & mixture & $\xi_1$ & 2.8997 & 0.5365 & & & \\
    & & $\xi_2$ & 0.3485 & 0.0850 & & & \\
    & Student-$t$ ($\nu = 8$) & $\xi_2$ & 1.2254 & 0.1191 & -2219.88 & \textbf{4587.65} & \textbf{4742.23} \\
    & exchangeable & & & & & & \\
    & Student-$t$ ($\nu = 8$) & $\pi$ & 0.2496 & 0.4007 & \textbf{-2219.67} & 4588.32 & 4743.08 \\
    & mixture & $\xi_1$ & 2.7921 & 1.3011 & & & \\
    & & $\xi_2$ & 0.9632 & 0.4314 & & & \\
    \hline
    Negative & Gaussian & $\xi_2$ & 0.9550 & 0.1141 & -1928.56 & 3889.84 & 3924.25 \\
    Binomial & exchangeable & & & & & & \\
    & Gaussian & $\pi$ & 0.6836 & 0.1174 & -1919.69 & 3873.65 & 3909.26 \\
    & mixture & $\xi_1$ & 1.4546 & 0.4483 & & & \\
    & & $\xi_2$ & 0.1195 & 0.0799 & & & \\
    & Student-$t$ ($\mathbf{\nu} = 21$) & $\xi_2$ & 0.9172 & 0.1168 & -1924.85 & 3881.29 & 3916.95 \\
    & exchangeable & & & & & & \\
    & Student-$t$ ($\mathbf{\nu} = 22$) & $\pi$ & 0.6976 & 0.1244 & \textbf{-1919.48} & \textbf{3872.90} & \textbf{3908.15} \\
    & mixture & $\xi_1$ & 1.3930 & 0.4360 & & & \\
    & & $\xi_2$ & 0.1261 & 0.0882 & & & \\
    \hline
    \end{tabular}}
    \end{small}
    \caption{Estimated dependence parameters and their standard errors for the epilepsy data, using both standard and mixture elliptical copulas. The maximum composite log-likelihood value, CLAIC, and CLBIC for each model are reported.}
    \label{tab:realdata2fit2}
\end{table}
\par
\begin{figure}
    \centering
    \includegraphics[width = 12cm]{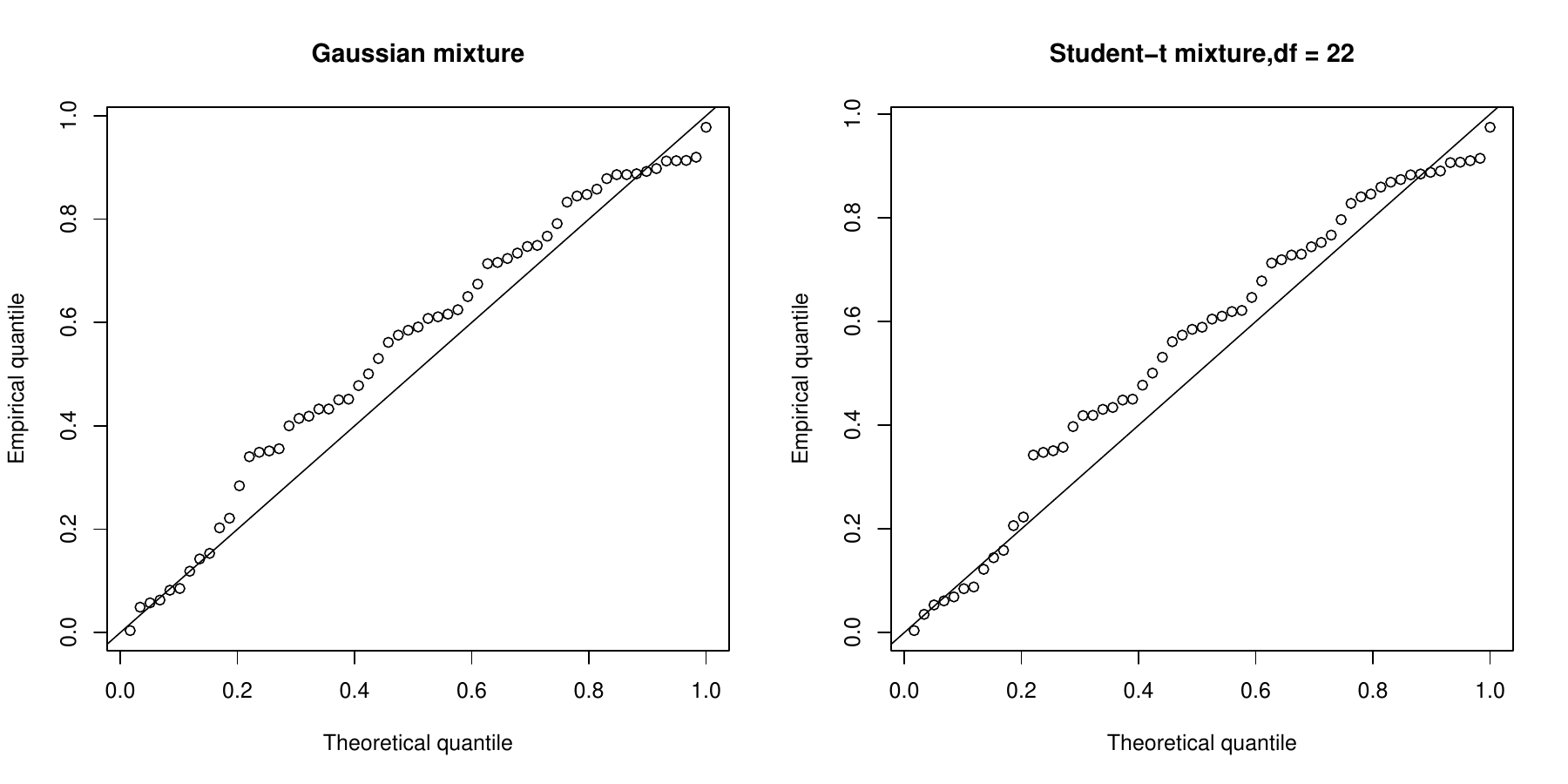}
    \caption{$t$-plots for the mixture of elliptical copulas for the epilepsy data.}
    \label{fig:data2.t-plot}
\end{figure}

 The obtained results are reported in Table \ref{tab:realdata2fit1} and \ref{tab:realdata2fit2}. As expected from this data set, Negative Binomial based models outperform the Poisson based models and the parameter $\psi$ captured the over dispersion present in the data. The estimate of $\beta_2$ is negative, which implies patients under the progabide treatment group has lower seizure counts compared to the control group. Based on the estimate of $\beta_3$, it shows a positive relation between age and seizure rate. The estimate of $\beta_1$ shows, the patients who initiated with high seizure rate continued with the same throughout. The estimated standard errors of the marginal parameters are comparably large, which is due to the fact that the sample size is quite low in this data set. Let's look now to the estimated dependence parameters of the Negative Binomial based models. Student-$t$ and Gaussian mixture copula are not significantly different for this data set based on the estimated value of the degrees of freedom parameter and the selection criteria. Interestingly, based on the estimate of $\pi$, it seems that AR(1) structure is more prominent, but as we see from relatively high value of $\xi_1$, it shows the first component copula is pretty close to the independence copula. As we did previously, we obtain the concordance matrices and compare with their empirical versions for the epilepsy data set. Let $A(\tau)$, $A(\hat{\tau})$, $A(\rho)$ and $A(\hat{\rho})$ be the matrices described in Sub-section \ref{subsec81} given as
\begin{align}
A(\tau) & = \left[\arraycolsep=3.0pt\def\arraystretch{1.0}\begin{array}{cccc} 1.00 \\ 0.34 & 1.00 \\ 0.21 & 0.28 & 1.00 \\ 0.26 & 0.40 & 0.28 & 1.00 \end{array}\right], A(\hat{\tau}) = \left[\arraycolsep=3.0pt\def\arraystretch{1.0}\begin{array}{cccc} 1.00 \\ 0.30 & 1.00 \\ 0.21 & 0.30 & 1.00 \\ 0.19 & 0.21 & 0.30 & 1.00 \end{array}\right], \nonumber \\ A(\rho) & = \left[\arraycolsep=3.0pt\def\arraystretch{1.0}\begin{array}{cccc} 1.00 \\ 0.47 & 1.00 \\ 0.30 & 0.39 & 1.00 \\ 0.37 & 0.53 & 0.40 & 1.00 \end{array}\right] \text{and} \; A(\hat{\rho}) = \left[\arraycolsep=3.0pt\def\arraystretch{1.0}\begin{array}{cccc} 1.00 \\ 0.43 & 1.00 \\ 0.30 & 0.42 & 1.00 \\ 0.27 & 0.30 & 0.42 & 1.00 \end{array}\right]. \nonumber
\end{align}
Though the sample size is relatively low for this data set but we see that the best fitting mixture copula closely captured the longitudinal correlation. Finally the modified $t$-plot for the epilepsy data set is given in Figure \ref{fig:data2.t-plot} which shows both of these elliptical mixture copulas captured the temporal dependency similarly. Overall, our models showed substantial improvements to capture the longitudinal correlation than their alternatives. Since we used different covariate setup to analyse these data sets, our parametric results are different than those in \citet{sutradhar2003overview} or \citet{jowaheer2002analysing} but the conclusions are somewhat similar. However, widely used random effect models can be difficult to interpret and sometimes poses computational challenges in-order to estimate the model parameters due to involvement of multidimensional integrations. Copula based models are in general good alternatives to those, since they can be estimated under efficient one-stage or two-stage estimation procedures with valid standard errors.
\section{Discussion}\label{sec9}
Modeling the correlation structure of discrete longitudinal data is a crucial area of research, particularly given the absence of suitable multivariate distributions. In recent years, the copula approach has emerged as a primary technique for modeling the dependence structure of multivariate data across various disciplines. However, its application in discrete data modeling has been relatively overlooked in the literature due to theoretical limitations. Drawing inspiration from the literature on finite mixture models, this article introduces the concept of a mixture of elliptical copulas to model the temporal dependence of longitudinal count data. We derive the dependence properties of finite mixture copulas for both continuous and discrete cases which is our key contribution. Under a full parametric setup, we estimate the parameters of our proposed class of models using a two-stage composite likelihood method and validate them through extensive simulation studies. Additionally, we modify and extend the $t$-plot method of model validation for elliptical copulas. By allowing different correlation structures in each copula component, we demonstrate that our mixture copula models offer better insights into the dependence structure of count longitudinal data. Although we focus on balanced cases in our analysis, the approach can be readily extended to unbalanced longitudinal data as long as mixture of elliptical copulas is considered. While we assume the number of mixing components to be known for simplicity in this article, it would be intriguing to extend our approach to cases where this parameter is unknown, as described by \citet{yang2022time}. One drawback of utilizing elliptical copulas is their tendency to capture symmetric dependence, which also holds true for their mixture. This means by construction the dependence measures are convex combinations of the component wise copulas. We address this issue in another work where we introduced geometric skew-normal copula constructed using multivariate geometric skew-normal distribution which is also closed under marginalization. Since we implemented two-stage procedure we only studied simulated models under correct specifications. It is an interesting exercise to monitor the performance of the parameter estimates under some misspecified models. In that setting Bayesian methods for one step estimation of the model parameters could be considered. We look forward to exploring these in our future research endeavors. \\
\\
\textbf{Acknowledgement}: The author expresses heartfelt gratitude to Prof. Brajendra C. Sutradhar from Memorial University in St. John's Canada for his thorough reading of an earlier version of the manuscript and for his valuable suggestions for improvements. The author also expresses his sincere greetings to the Editor in Chief of METRON and two anonymous referees whose constructive comments led to an improved version of the manuscript.
\\ \\
\textbf{Availability of programs:} The R code used to conduct this study is available upon request from the author.
\\ \\
\textbf{Conflict of interest:} There is no conflict of interest from the author.
\bibliographystyle{agsm}
\bibliography{paper5ref1}
\appendix
\section{Appendix}\label{appndx1}
Here we present the proofs of the theorems stated in Section \ref{sec3}. \\ \\
\textbf{Proof of Theorem \ref{thm1}:}
Let $(X_1',X_2')^\intercal$ be an independent copy of $(X_1,X_2)^\intercal$. Using the definition of Kendall's tau we have -
\begin{align}\label{kend1der1}
\tau(C_{\text{mix}}) & = P(\text{concordance}) - P(\text{discordance}) \nonumber \\ & = P((X_1 - X_1')(X_2 - X_2') > 0) - P((X_1 - X_1')(X_2 - X_2') < 0) \nonumber \\ & = 4P(X_1' < X_1,X_2' < X_2) - 1 \nonumber \\ & = 4\int_0^1\int_0^1 C_{\text{mix}}(u_1,u_2) dC_{\text{mix}}(u_1,u_2) - 1.
\end{align}
\begin{align}
& \text{Therefore}, \quad \int_0^1\int_0^1 C_{\text{mix}}(u_1,u_2) dC_{\text{mix}}(u_1,u_2) \nonumber \\ & = \int_0^1\int_0^1 \sum_{l=1}^K \pi_l C_l(u_1,u_2|\mathbf{\phi}_l) d\sum_{l=1}^K \pi_l C_l(u_1,u_2|\mathbf{\phi}_l) \nonumber \\ & = \sum_{l=1}^K \pi^2_l \int_0^1\int_0^1 C_l(u_1,u_2|\mathbf{\phi}_l) dC_l(u_1,u_2|\mathbf{\phi}_l) + \sum_{l\neq m}^K \pi_l\pi_m \int_0^1\int_0^1 C_l(u_1,u_2|\mathbf{\phi}_l) dC_m(u_1,u_2|\mathbf{\phi}_m). \nonumber \\
& \text{Now}, \quad \int_0^1\int_0^1 C_l(u_1,u_2|\mathbf{\phi}_l) dC_m(u_1,u_2|\mathbf{\phi}_m) \nonumber \\ & = \int_0^1\int_0^1 \int_0^{u_1}\int_0^{u_2} c_l(u_1,u_2|\mathbf{\phi}_l) c_m(u_1,u_2|\mathbf{\phi}_m) du_1^2 du_2^2 \nonumber \\ & = \int_0^1\int_0^1 C_m(u_1,u_2|\mathbf{\phi}_m) dC_l(u_1,u_2|\mathbf{\phi}_l). \nonumber 
\end{align}
Since $\sum_{l=1}^K \pi_l = 1$, by plugging the values in equation (\ref{kend1der1}) we obtain the result. \\ \\
\textbf{Proof of Theorem \ref{thm2}:}
Let $(X_1^*,X_2^*)^\intercal$ be two independent random variables (i.e. bivariate random vector with independence copula) with same marginal distributions $F_j, j = 1,2$. Using the definition of Spearman's rho we have -
\begin{align}\label{sper1der1}
\rho(C_{\text{mix}}) & = 3(P(\text{concordance}) - P(\text{discordance})) \nonumber \\ & = 3(P((X_1 - X_1^*)(X_2 - X_2^*) > 0) - P((X_1 - X_1^*)(X_2 - X_2^*) < 0)) \nonumber \\ & = 12P(X_1^* < X_1,X_2^* < X_2) - 3 \nonumber \\ & = 12\int_0^1\int_0^1 C_{\text{mix}}(u_1,u_2) du_1 du_2 - 3.
\end{align}
\begin{align}
& \text{Therefore}, \quad \int_0^1\int_0^1 C_{\text{mix}}(u_1,u_2) du_1 du_2 \nonumber \\ & = \int_0^1\int_0^1 \sum_{l=1}^K \pi_l C_l(u_1,u_2|\mathbf{\phi}_l) du_1 du_2 = \sum_{l=1}^K \pi_l \int_0^1\int_0^1 C_l(u_1,u_2|\mathbf{\phi}_l) du_1 du_2. \nonumber
\end{align}
Since $\sum_{l=1}^K \pi_l = 1$, by plugging the values in equation (\ref{sper1der1}) we obtain the result. \\ \\
\textbf{Proof of Theorem \ref{thm3}:}
When $(X_1,X_2)^\intercal$ is integer valued random vector the probability of tie is non-zero and $P(\text{concordance}) + P(\text{discordance}) + P(\text{tie}) = 1$. Therefore we have -
\begin{align}\label{kend1der2}
\tau^*(C_{\text{mix}}) & = P(\text{concordance}) - P(\text{discordance}) \nonumber \\ & = 2P(\text{concordance}) + P(\text{tie}) - 1 \nonumber \\ & = 2P((X_1 - X_1')(X_2 - X_2') > 0) + P(X_1 = X_1' \cup X_2 = X_2') - 1 \nonumber \\ & = 4P(X_1' < X_1,X_2' < X_2) + P(X_1 = X_1' \cup X_2 = X_2') - 1.
\end{align}
The last expression is due to the fact that $(X_1',X_2')^\intercal$ and $(X_1,X_2)^\intercal$ are identically distributed.
\begin{align}\label{kendir3}
& \text{Now}, \quad P(X_1' < X_1,X_2' < X_2) \nonumber \\ & = \sum_{x_1=0}^\infty \sum_{x_2=0}^\infty P(X_1' \leq x_1-1,X_2' \leq x_2-1) P(X_1 = x_1,X_2 = x_2) \nonumber \\ & = \sum_{x_1=0}^\infty \sum_{x_2=0}^\infty C_{\text{mix}}(F_1(x_1-1),F_2(x_2-1)) h_{\text{mix}}(x_1,x_2) \nonumber \\ & = \sum_{x_1=0}^\infty \sum_{x_2=0}^\infty \Big(\sum_{l=1}^K \pi_l C_l(F_1(x_1-1),F_2(x_2-1)|\mathbf{\phi}_l)\Big) \Big(\sum_{l=1}^K \pi_l h_l(x_1,x_2)\Big) \nonumber \\ & = \sum_{l=1}^K \pi^2_l \sum_{x_1=0}^\infty \sum_{x_2=0}^\infty C_l(F_1(x_1-1),F_2(x_2-1)|\mathbf{\phi}_l) h_l(x_1,x_2) \nonumber \\ & + \sum_{l\neq m}^K \pi_l\pi_m \sum_{x_1=0}^\infty \sum_{x_2=0}^\infty C_l(F_1(x_1-1),F_2(x_2-1)|\mathbf{\phi}_l) h_m(x_1,x_2).
\end{align}
\begin{align}\label{kendir4}
& \text{And}, \quad P(X_1 = X_1' \cup X_2 = X_2') \nonumber \\ & = P(X_1 = X_1') + P(X_2 = X_2') - P(X_1 = X_1',X_2 = X_2') \nonumber \\ & = \sum_{x_1=0}^\infty f_1^2(x_1) + \sum_{x_2=0}^\infty f_2^2(x_2) - \sum_{x_1=0}^\infty \sum_{x_2=0}^\infty h_{\text{mix}}^2(x_1,x_2) \nonumber \\ & = \sum_{x_1=0}^\infty f_1^2(x_1) + \sum_{x_2=0}^\infty f_2^2(x_2) - \sum_{l=1}^K \pi^2_l \sum_{x_1=0}^\infty \sum_{x_2=0}^\infty h_l^2(x_1,x_2) \nonumber \\ & - \sum_{l\neq m}^K \pi_l\pi_m \sum_{x_1=0}^\infty \sum_{x_2=0}^\infty h_l(x_1,x_2)h_m(x_1,x_2).
\end{align}
\begin{align}
\text{Let}, \quad \tau^*(C_l) & = \sum_{x_1=0}^\infty \sum_{x_2=0}^\infty h_l(x_1,x_2)\{4C_l(F_1(x_1-1),F_2(x_2-1)|\mathbf{\phi}_l) - h_l(x_1,x_2)\} \nonumber \\
& + \sum_{x_1=0}^\infty f_1^2(x_1) + \sum_{x_2=0}^\infty f_2^2(x_2) - 1, \nonumber \\ Q^*_{lm} & = \sum_{x_1=0}^\infty \sum_{x_2=0}^\infty h_m(x_1,x_2)\{4C_l(F_1(x_1-1),F_2(x_2-1)|\mathbf{\phi}_l) - h_l(x_1,x_2)\} \nonumber \\
& + \sum_{x_1=0}^\infty f_1^2(x_1) + \sum_{x_2=0}^\infty f_2^2(x_2) - 1. \nonumber
\end{align}
Therefore, using (\ref{kendir3}) and (\ref{kendir4}) in (\ref{kend1der2}) we get -
\begin{align}
\tau^*(C_{\text{mix}}) & = \sum_{l=1}^K \pi^2_l \Big(\tau^*(C_l) + \sum_{x_1=0}^\infty \sum_{x_2=0}^\infty h_l^2(x_1,x_2) - \sum_{x_1=0}^\infty f_1^2(x_1) - \sum_{x_2=0}^\infty f_2^2(x_2) + 1\Big) \nonumber \\ & + \sum_{l\neq m}^K \pi_l\pi_m \Big(Q^*_{lm} + \sum_{x_1=0}^\infty \sum_{x_2=0}^\infty h_l(x_1,x_2)h_m(x_1,x_2) - \sum_{x_1=0}^\infty f_1^2(x_1) - \sum_{x_2=0}^\infty f_2^2(x_2) + 1\Big) \nonumber \\ & - \sum_{l=1}^K \pi^2_l \sum_{x_1=0}^\infty \sum_{x_2=0}^\infty h_l^2(x_1,x_2) - \sum_{l\neq m}^K \pi_l\pi_m \sum_{x_1=0}^\infty \sum_{x_2=0}^\infty h_l(x_1,x_2)h_m(x_1,x_2) \nonumber \\ & + \sum_{x_1=0}^\infty f_1^2(x_1) + \sum_{x_2=0}^\infty f_2^2(x_2) - 1 \nonumber \\ & = \sum_{l=1}^K \pi^2_l \tau^*(C_l) + \sum_{l\neq m}^K \pi_l\pi_m Q^*_{lm}, \nonumber
\end{align}
since $\sum_{l=1}^K \pi^2_l + \sum_{l\neq m}^K \pi_l\pi_m - 1 = 0$, and hence the proof is completed. \\ \\
\textbf{Proof of Theorem \ref{thm4}:}
Using the same definition we have -
\begin{align}\label{rhodir1}
\rho^*(C_{\text{mix}}) & = 3(P(\text{concordance}) - P(\text{discordance})) \nonumber \\ & = 3(2P(\text{concordance}) + P(\text{tie}) - 1) \nonumber \\ & = 6P((X_1 - X_1^*)(X_2 - X_2^*) > 0) + 3(P(X_1 = X_1^* \cup X_2 = X_2^*) - 1) \nonumber \\ & = 6P(X_1^* < X_1,X_2^* < X_2) + 6P(X_1^* > X_1,X_2^* > X_2) \nonumber \\ & + 3(P(X_1 = X_1^* \cup X_2 = X_2^*) - 1).
\end{align}
The last expression is due to the fact that $(X_1^*,X_2^*)^\intercal$ and $(X_1,X_2)^\intercal$ have different joint distribution.
\begin{align}\label{rhodir2}
& \text{Now}, \quad P(X_1^* < X_1,X_2^* < X_2) \nonumber \\ & = \sum_{x_1=0}^\infty \sum_{x_2=0}^\infty P(X_1^* \leq x_1-1,X_2^* \leq x_2-1) P(X_1 = x_1,X_2 = x_2) \nonumber \\ & = \sum_{x_1=0}^\infty \sum_{x_2=0}^\infty F_1(x_1-1)F_2(x_2-1) h_{\text{mix}}(x_1,x_2) \nonumber \\ & = \sum_{x_1=0}^\infty \sum_{x_2=0}^\infty F_1(x_1-1)F_2(x_2-1) \Big(\sum_{l=1}^K \pi_l h_l(x_1,x_2)\Big) \nonumber \\ & = \sum_{l=1}^K \pi_l \sum_{x_1=0}^\infty \sum_{x_2=0}^\infty F_1(x_1-1)F_2(x_2-1) h_l(x_1,x_2).
\end{align}
\begin{align}\label{rhodir3}
& \text{And}, \quad P(X_1^* > X_1,X_2^* > X_2) \nonumber \\ & = \sum_{x_1=0}^\infty \sum_{x_2=0}^\infty P(X_1^* > x_1,X_2^* > x_2-1) P(X_1 = x_1,X_2 = x_2) \nonumber \\ & = \sum_{x_1=0}^\infty \sum_{x_2=0}^\infty (1 - F_1(x_1))(1 - F_2(x_2)) h_{\text{mix}}(x_1,x_2) \nonumber \\ & = \sum_{x_1=0}^\infty \sum_{x_2=0}^\infty (1 - F_1(x_1))(1 - F_2(x_2)) \Big(\sum_{l=1}^K \pi_l h_l(x_1,x_2)\Big) \nonumber \\ & = \sum_{l=1}^K \pi_l \sum_{x_1=0}^\infty \sum_{x_2=0}^\infty (1 - F_1(x_1))(1 - F_2(x_2)) h_l(x_1,x_2).
\end{align}
\begin{align}\label{rhodir4}
& \text{And}, P(X_1 = X_1^* \cup X_2 = X_2^*) \nonumber \\ & = P(X_1 = X_1^*) + P(X_2 = X_2^*) - P(X_1 = X_1^*,X_2 = X_2^*) \nonumber \\ & = \sum_{x_1=0}^\infty f_1^2(x_1) + \sum_{x_2=0}^\infty f_2^2(x_2) - \sum_{x_1=0}^\infty \sum_{x_2=0}^\infty f_1(x_1)f_2(x_2) h_{\text{mix}}(x_1,x_2) \nonumber \\ & = \sum_{x_1=0}^\infty f_1^2(x_1) + \sum_{x_2=0}^\infty f_2^2(x_2) - \sum_{l=1}^K \pi_l \sum_{x_1=0}^\infty \sum_{x_2=0}^\infty f_1(x_1)f_2(x_2) h_l(x_1,x_2).
\end{align}
\begin{align}
\text{Let}, \quad \rho^*(C_l) & = \sum_{x_1=0}^\infty \sum_{x_2=0}^\infty h_l(x_1,x_2) \{6F_1(x_1-1)F_2(x_2-1) + 6(1 - F_1(x_1))(1 - F_2(x_2)) \nonumber \\ & - 3f_1(x_1)f_2(x_2)\} + 3(\sum_{x_1=0}^\infty f_1^2(x_1) + \sum_{x_2=0}^\infty f_2^2(x_2) - 1). \nonumber
\end{align}
Therefore, using (\ref{rhodir2}), (\ref{rhodir3}) and (\ref{rhodir4}) in (\ref{rhodir1}) we get - 
\begin{align}
\rho^*(C_{\text{mix}}) & = \sum_{l=1}^K \pi_l \Big(\rho^*(C_l) + 3\sum_{x_1=0}^\infty \sum_{x_2=0}^\infty f_1(x_1)f_2(x_2) h_l(x_1,x_2) - 3\Big(\sum_{x_1=0}^\infty f_1^2(x_1) + \sum_{x_2=0}^\infty f_2^2(x_2) - 1\Big) \nonumber \\ & + 3\Big(\sum_{x_1=0}^\infty f_1^2(x_1) + \sum_{x_2=0}^\infty f_2^2(x_2) - 1 - \sum_{l=1}^K \pi_l \sum_{x_1=0}^\infty \sum_{x_2=0}^\infty f_1(x_1)f_2(x_2) h_l(x_1,x_2)\Big) \nonumber \\ & = \sum_{l=1}^K \pi_l \rho^*(C_l), \nonumber
\end{align}
and hence the proof is completed. \\ \\
\textbf{Proof of Theorem \ref{thm5}:}
Straight from the definition we have -
\begin{align}
\lambda_U(C_{\text{mix}}) & = \lim_{u \to 1-} \frac{1 -2u + C_{\text{mix}}(u,u)}{1-u} = \lim_{u \to 1-} \frac{1 -2u + \sum_{l=1}^K \pi_l C_l(u,u)}{1-u} \nonumber \\ & = \sum_{l=1}^K \pi_l \lim_{u \to 1-} \frac{1 -2u + C_l(u,u)}{1-u} = \sum_{l=1}^K \pi_l \lambda_U(C_l). \nonumber
\end{align}
\begin{align}
\lambda_L(C_{\text{mix}}) & = \lim_{u \to 0+} \frac{C_{\text{mix}}(u,u)}{u} = \lim_{u \to 0+} \frac{\sum_{l=1}^K \pi_l C_l(u,u)}{u} \nonumber \\ & = \sum_{l=1}^K \pi_l \lim_{u \to 0+} \frac{C_l(u,u)}{u} = \sum_{l=1}^K \pi_l \lambda_L(C_l). \nonumber
\end{align}
\end{document}